\documentclass[10pt,prd,preprintnumbers,amsmath,amssymb,nofootinbib]{revtex4}

\usepackage{graphics}
\usepackage{bm}
\usepackage{cancel}
\usepackage{color}
\usepackage{graphicx}
\usepackage{multirow}
\usepackage{amsmath}
\usepackage{slashed}

\usepackage{array}                                                   

\makeatletter
\newcommand{\thickhline}{\noalign {\ifnum 0=`}\fi \hrule height 1pt\futurelet \reserved@a \@xhline}
\newcolumntype{"}{@{\hskip\tabcolsep\vrule width 1pt\hskip\tabcolsep}}                             
\makeatother

\usepackage{graphicx} 

\usepackage[colorlinks,
            citecolor=blue,
            anchorcolor=red,
            menucolor=red,
            linkcolor=red,
            filecolor=red,
            runcolor=red,
            urlcolor=blue,
            frenchlinks=red]{hyperref}

\begin{document}

\title{Pentaquark states with the $QQQq\bar{q}$ configuration in a simple model}

\author{Shi-Yuan Li$^1$, Yan-Rui Liu$^1$, Yu-Nan Liu$^1$, Zong-Guo Si$^{1,2}$, Jing Wu$^1$}
\affiliation{$^1$School of Physics, Shandong University, Jinan 250100, P. R. China}
\affiliation{$^2$Institute of Theoretical Physics, Chinese Academy of Science, Beijing 100190, P. R. China}

\date{\today}
\begin{abstract}
We discuss the mass splittings for the $S$-wave triply heavy pentaquark states with the $QQQq\bar{q}$ $(Q=b,c;q=u,d,s)$ configuration which is a mirror structure of $Q\bar{Q}qqq$. The latter configuration is related with the nature of $P_c(4380)$ observed by the LHCb Collaboration. The considered pentaquark masses are estimated with a simple method. One finds that such states are probably not narrow even if they do exist. This leaves room for molecule interpretation for a state around the low-lying threshold of a doubly heavy baryon and a heavy-light meson, e.g. $\Xi_{cc}D$, if it were observed. As a by product, we conjecture that upper limits for the masses of the conventional triply heavy baryons can be determined by the masses of the conventional doubly heavy baryons.
\end{abstract}
\maketitle

\section{INTRODUCTION}\label{sec1}

Recently, the LHCb Collaboration confirmed the $\Xi_{cc}$ baryon \cite{Aaij:2017ueg} which was predicted in the quark model (QM) decades ago \cite{DeRujula:1975qlm} and first measured by the SELEX Collaboration in 2002 \cite{Mattson:2002vu}. Excited conventional heavy hadrons are also observed in recent years \cite{Chen:2016spr}. We are still on the way to confirm the conventional structure with three heavy quarks. However, the most exciting thing in hadron physics is that more and more unexpected hadrons are detected \cite{Liu:2013waa,Esposito:2014rxa,Olsen:2014qna,Hosaka:2016pey,Chen:2016qju,Richard:2016eis,Lebed:2016hpi,Esposito:2016noz,Guo:2017jvc,Ali:2017jda,Olsen:2017bmm,Karliner:2017qhf} and thus hadrons beyond the conventional quark model should exist. This indicates that we may identify the existence of multiquark states before the confirmation of all the QM states. For example, the $P_c(4380)$ and $P_c(4450)$ states observed in the $J/\psi N$ channel by the LHCb Collaboration \cite{Aaij:2015tga} can be interpreted as pentaquark states since their high masses are not natural for conventional 3q baryons \cite{Wu:2010jy}. If a pentaquark can be confirmed, the existence of other multiquark states should also be possible. It is interesting to investigate theoretically which multiquark systems allow bound states.

Since the pentaquark-like states $P_c(4380)$ and $P_c(4450)$ are much higher than the $J/\psi N$ threshold and the $J/\psi N$ interaction is not strong (a lattice simulation gives a value around 0.7 fm for the scattering length \cite{Yokokawa:2006td}), the interpretation with $J/\psi N$ scattering states seem not to be appropriate. In the literature, various interpretations, such as $\Sigma_c^{(*)}\bar{D}^*$ molecules, diquark-diquark-antiquark pentaquarks or diquark-triquark pentaquarks, have been used to understand their nature \cite{Chen:2016qju,Richard:2016eis,Lebed:2016hpi,Esposito:2016noz,Guo:2017jvc,Ali:2017jda,Olsen:2017bmm,Karliner:2017qhf}. Assigning them as $qqqc\bar{c}$ compact pentaquarks is another possible interpretation, where the $c\bar{c}$ is a color-octet state. In Refs. \cite{Takeuchi:2016ejt,Wu:2017weo,Li:2017ghe}, investigations along this line were performed. It was found that interpreting the $P_c(4380)$ state as such a pentaquark can be accepted. Further investigations on this type configuration of pentaquarks showed that a stable $\Lambda$-type hidden-charm state is also possible \cite{Wu:2017weo,Irie:2017qai}. In the present study, we are going to consider a mirror-type structure by changing heavy (light) quarks to light (heavy) quarks, pentquarks with the $QQQq\bar{q}$ configuration where $QQQ$ is always a color-octet triply heavy state. In the following discussions, we use $QQQ$ to denote $ccc$, $bbb$, $ccb$, or $bbc$. When one needs to distinguish the quark contents, we will also use $QQQ$ to denote $ccc$ or $bbb$ and $QQQ'$ to denote $ccb$ or $bbc$.

In the study of the $udsQ\bar{Q}$ states with a colored $uds$, one found that the quark-quark and quark-antiquark color-spin interactions can all be attractive \cite{Wu:2017weo}. Mixing effects among different color-spin structures further lowered the pentaquark masses and finally resulted in a $\Lambda$-type state which has a rather low mass. The mass is not far from the value obtained in Ref. \cite{Wu:2010jy} where molecule configuration was adopted and hidden-charm pentaquarks were proposed first. In the present work, we try to understand whether colored $QQQ$ is also helpful to the formation of pentqaurks or not. If the answer is yes, a problem to distinguish the pentaquarks from conventional $QQQ$ baryons may arise. If no, one expects that conventional triply heavy baryons rather than pentaquarks would be found experimentally first.

Up to now, theoretical studies have given the masses of the conventional $QQQ$ baryons in various approaches, although there is still no experimental evidence about them. According to these calculations, the mass of the ground $\Omega_{ccc}$ baryon, for example, is in the range of 4.6$\sim$5.0 GeV \cite{Chen:2016spr,Vijande:2004at,Wang:2011ae,Wei:2015gsa,Weng:2018mmf}. If an additional quark-antiquark pair surrounds the heavy color source, there is a possibility that the system has a little higher mass and it looks like an excited $\Omega_{ccc}$ state. This possibility would be excluded if the $QQQq\bar{q}$ state has a much higher mass. We would like to extract some information about this problem from the following investigations.

In this article, we consider the mass splittings of the compact triply heavy pentaquarks $QQQq\bar{q}$ $(Q=b,c;q=u,d,s)$\footnote{Ref. \cite{Wang:2018ihk} presents a study of the triply heavy pentaquarks with the $QQ\bar{Q}qq$ configuration.} and estimate their rough masses in the framework of a simple quark model. In the following Sec. \ref{sec2}, the wave functions will be constructed. In Sec. \ref{sec3}, we give the chromomagnetic interaction (CMI) matrices for different types of pentaquarks. In Sec. \ref{sec4}, we show our choice of relevant parameters and analyse the numerical results for the spectrum. The last section is a short summary.

\section{Construction of the wave functions}\label{sec2}

Before the calculation of CMI matrices for the considered ground pentaquark states, we need to construct all the spin-color wave functions. The configuration we consider contains three heavy quarks ($ccc$, $ccb$, $bbc$, or $bbb$) and a light quark-antiquark pair ($n\bar{n}$, $n\bar{s}$, $s\bar{n}$, or $s\bar{s}$, $n=u$ or $d$), which is simply denoted as $1234\bar{5}$. Table \ref{Flavor} lists the flavor contents of these 16 systems.
\begin{table}[htb]
\caption{Flavor contents of the pentaquarks we consider.}\label{Flavor}
\centering
\begin{tabular}{|c|c|c|c|c|}
\hline
$|1234\bar{5}\rangle_f$ & $|4\bar{5}\rangle_f=|n\bar{n}\rangle$ & $|4\bar{5}\rangle_f=|n\bar{s}\rangle$ & $|4\bar{5}\rangle_f=|s\bar{n}\rangle$ & $|4\bar{5}\rangle_f=|s\bar{s}\rangle$ \\
 \hline
$|123\rangle_f=|ccc\rangle$ & $|cccn\bar{n}\rangle$ & $|cccn\bar{s}\rangle$ & $|cccs\bar{n}\rangle$ & $|cccs\bar{s}\rangle$ \\
 \hline
$|123\rangle_f=|ccb\rangle$ & $|ccbn\bar{n}\rangle$ & $|ccbn\bar{s}\rangle$ & $|ccbs\bar{n}\rangle$ & $|ccbs\bar{s}\rangle$ \\
 \hline
$|123\rangle_f=|bbc\rangle$ & $|bbcn\bar{n}\rangle$ & $|bbcn\bar{s}\rangle$ & $|bbcs\bar{n}\rangle$ & $|bbcs\bar{s}\rangle$ \\
 \hline
$|123\rangle_f=|bbb\rangle$ & $|bbbn\bar{n}\rangle$ & $|bbbn\bar{s}\rangle$ & $|bbbs\bar{n}\rangle$ & $|bbbs\bar{s}\rangle$ \\
\hline
\end{tabular}
\end{table}

Here, we use $|S_{12},S_{123},S_{4\bar{5}},J=S_{1234\bar{5}}\rangle$ to denote the possible spin wave functions. There are five states with $J=1/2$,
\begin{eqnarray}
X_1=|1,\frac12,0,\frac12\rangle, \quad X_2=|1,\frac12,1,\frac12\rangle, \quad X_3=|1,\frac32,1,\frac12\rangle,\quad
X_4=|0,\frac12,1,\frac12\rangle, \quad X_5=|0,\frac12,0,\frac12\rangle,
\end{eqnarray}
four states with $J=3/2$,
\begin{eqnarray}
X_6=|1,\frac12,1,\frac32\rangle,\quad X_7=|1,\frac32,0,\frac32\rangle,\quad X_8=|1,\frac32,1,\frac32\rangle,\quad X_9=|0,\frac12,1,\frac32\rangle,
\end{eqnarray}
and one state with $J=5/2$
\begin{eqnarray}
X_{10}=|1,\frac32,1,\frac52\rangle.
\end{eqnarray}
Their explicit expressions are easy to get by using the $SU(2)$ C.G. coefficients.

In color space, one uses $|R_{12},R_{123},R_{4\bar{5}},R_{1234\bar{5}}=1_c\rangle$ to denote the wave functions. Then, we find two bases
\begin{eqnarray}
 C_1&=&|6,8^{MS},8,1^{MS}\rangle,\quad
 C_2=|\bar{3},8^{MA},8,1^{MA}\rangle
\end{eqnarray}
for the present investigation, where the superscripts $MS$ and $MA$ mean that the first two quarks are symmetric and antisymmetric, respectively. Their explicit expressions are the same as those presented in Eq. (2) of Ref. \cite{Wu:2017weo}.

Taking the Pauli principle into account when we combine the bases in different spaces, one may obtain five types of total wave functions.
\begin{itemize}
\item Type A [Flavor$=QQQq\bar{q}$, J=1/2]:
\begin{eqnarray}\label{Awavefunction}
\Phi_1^A&=&\frac{1}{\sqrt{2}}\left\{[(QQ)_6^0Q]_8^{\frac{1}{2}}(q\bar{q})_8^0\right\}_1^{\frac{1}{2}}-
         \frac{1}{\sqrt{2}}\left\{[(QQ)_{\bar{3}}^1Q]_8^{\frac{1}{2}}(q\bar{q})_8^0\right\}_1^{\frac{1}{2}}
         =\frac{1}{\sqrt{2}}QQQq\bar{q}\otimes(C_1\otimes X_5-C_2\otimes X_1),\nonumber\\
\Phi_2^A&=&\frac{1}{\sqrt{2}}\left\{[(QQ)_6^0Q]_8^{\frac{1}{2}}(q\bar{q})_8^1\right\}_1^{\frac{1}{2}}-
         \frac{1}{\sqrt{2}}\left\{[(QQ)_{\bar{3}}^1Q]_8^{\frac{1}{2}}(q\bar{q})_8^1\right\}_1^{\frac{1}{2}}
         =\frac{1}{\sqrt{2}}QQQq\bar{q}\otimes(C_1\otimes X_4-C_2\otimes X_2);
\end{eqnarray}
\item Type B [Flavor$=QQQq\bar{q}$, J=3/2]:
\begin{equation}\label{Bwavefunction}
\Phi_1^B=\frac{1}{\sqrt{2}}\left\{[(QQ)_6^0Q]_8^{\frac{1}{2}}(q\bar{q})_8^1\right\}_1^{\frac{3}{2}}-
         \frac{1}{\sqrt{2}}\left\{[(QQ)_{\bar{3}}^1Q]_8^{\frac{1}{2}}(q\bar{q})_8^1\right\}_1^{\frac{3}{2}}
         =\frac{1}{\sqrt{2}}QQQq\bar{q}\otimes(C_1\otimes X_9-C_2\otimes X_6);
\end{equation}
\item Type C [Flavor$=QQQ'q\bar{q}$, J=1/2]:
\begin{eqnarray}\label{Cwavefunction}
 \Phi_1^C&=&\left\{[(QQ)_6^0Q']_8^{\frac{1}{2}}(q\bar{q})_8^0\right\}_1^{\frac{1}{2}}=(QQQ'q\bar{q})\otimes C_1\otimes X_5,\quad
\Phi_2^C=\left\{[(QQ)_6^0Q']_8^{\frac{1}{2}}(q\bar{q})_8^1\right\}_1^{\frac{1}{2}}=(QQQ'q\bar{q})\otimes C_1\otimes X_4,\nonumber\\
 \Phi_3^C&=&\left\{[(QQ)_{\bar{3}}^1Q']_8^{\frac{1}{2}}(q\bar{q})_8^0\right\}_1^{\frac{1}{2}}=(QQQ'q\bar{q})\otimes C_2\otimes X_1,\quad
\Phi_4^C=\left\{[(QQ)_{\bar{3}}^1Q']_8^{\frac{1}{2}}(q\bar{q})_8^1\right\}_1^{\frac{1}{2}}=(QQQ'q\bar{q})\otimes C_2\otimes X_2,\nonumber\\
\Phi_5^C&=&\left\{[(QQ)_{\bar{3}}^1Q']_8^{\frac{3}{2}}(q\bar{q})_8^1\right\}_1^{\frac{1}{2}}=(QQQ'q\bar{q})\otimes C_2\otimes X_3;
\end{eqnarray}
\item Type D [Flavor$=QQQ'q\bar{q}$, J=3/2]:
\begin{eqnarray}\label{Dwavefunction}
\Phi_1^D&=&\left\{[(QQ)_6^0Q']_8^{\frac{1}{2}}(q\bar{q})_8^1\right\}_1^{\frac{3}{2}}=(QQQ'q\bar{q})\otimes C_1\otimes X_9,\quad
\Phi_2^D=\left\{[(QQ)_{\bar{3}}^1Q']_8^{\frac{1}{2}}(q\bar{q})_8^1\right\}_1^{\frac{3}{2}}=(QQQ'q\bar{q})\otimes C_2\otimes X_6,\nonumber\\
\Phi_3^D&=&\left\{[(QQ)_{\bar{3}}^1Q']_8^{\frac{3}{2}}(q\bar{q})_8^0\right\}_1^{\frac{3}{2}}=(QQQ'q\bar{q})\otimes C_2\otimes X_7,\quad
\Phi_4^D=\left\{[(QQ)_{\bar{3}}^1Q']_8^{\frac{3}{2}}(q\bar{q})_8^1\right\}_1^{\frac{3}{2}}=(QQQ'q\bar{q})\otimes C_2\otimes X_8;\nonumber\\
\end{eqnarray}
\item Type E [Flavor$=QQQ'q\bar{q}$, J=5/2]:
\begin{eqnarray}\label{Ewavefunction}
\Phi_1^E=\left\{[(QQ)_{\bar{3}}^1Q']_8^{\frac{3}{2}}(q\bar{q})_8^1\right\}_1^{\frac{5}{2}}=(QQQ'q\bar{q})\otimes C_2\otimes X_{10}.
\end{eqnarray}
\end{itemize}
Here, $QQQ$ means $ccc$ or $bbb$ and $QQQ'$ means $ccb$ or $bbc$. The superscripts (subscripts) for quarks indicate spins (representations in color space). In each type of pentaquark systems, configuration mixing induced by the chromomagnetc interaction occurs. If we use $\Psi^X$ to denote a mixed wave function for the type-$X$ state, it can be written as a superposition of different configurations,
\begin{equation}\label{pentaquark's wave function}
\Psi^X=\sum_{i=1}^{N_X}C_i^X\Phi_i^X,
\end{equation}
where the coefficients $C_i^X$ satisfy the normalization condition $\sum_{i=1}^{N_X}|C_i^X|^2=1$ and $N_X$=2, 1, 5, 4, and 1 correspond to $X=A$, $B$, $C$, $D$, and $E$, respectively. There are $N_X$ independent $\Psi^X$.

\section{Chromomagnetic Interaction}\label{sec3}

The Hamiltonian for the mass calculation in the model reads
\begin{equation}\label{CMI}
\widehat{H}=\sum_{i=1}^nm_i-\sum_{i<j}\sum_{\alpha=1}^{3}\sum_{\beta=1}^{8}
C_{ij}(\sigma_i^{\alpha}\sigma_j^{\alpha})(\widetilde{\lambda}_i^{\beta}\widetilde{\lambda}_j^{\beta})
=\sum_{i=1}^nm_i+\widehat{H}_{CMI}.
\end{equation}
Here, $n$ is the number of (anti)quarks in the hadron and $m_i$ is the effective quark mass for the $i$th quark by taking account of effects from kinetic energy, color confinement, and so on. The coefficient $C_{ij}$ reflects the strength of the chromomagnetic interaction between the $i$th and $j$th quark components and is influenced by their masses. The Pauli matrix $\sigma_i^{\alpha}$ and Gell-Mann matrix $\widetilde{\lambda}_i^{\beta}$=$\lambda_i^{\beta}$ ($-{{\lambda}^*}_i^{\beta}$) act on the spin and color wave functions of the $i$th quark (antiquark), respectively.

With the constructed bases of wave functions for the type-$X$ pentaquarks, one can easily obtain the matrix element
$\textbf{[}H_{CMI}^X\textbf{]}_{kl}=\langle\Phi_k^X|{\widehat{H}}_{CMI}|\Phi_l^X\rangle$,
where $k$, $l$=1,2,...$N_X$. The calculation of all matrix elements gives five CMI matrices.
\begin{itemize}
\item Type A [Flavor$=QQQq\bar{q}$, J=1/2]:
\begin{eqnarray}\label{CMI1}
{{H}}_{CMI}^A=
\left(\begin{matrix}
10C_{12}+2C_{45} & \frac{10}{\sqrt{3}}(C_{14}+C_{15}) \\
 & 10C_{12}-\frac23 C_{45}-\frac{20}{3}(C_{14}-C_{15})
\end{matrix}\right).
\end{eqnarray}
Here, the base vector is ($\Phi_1^A$, $\Phi_2^A$).
\item Type B [Flavor$=QQQq\bar{q}$, J=3/2]:
\begin{eqnarray}\label{CMI2}
{{H}}_{CMI}^B=
10C_{12}-\frac23 C_{45}+\frac{10}{3}(C_{14}-C_{15}),
\end{eqnarray}
where the only base is $\Phi_1^B$.
\item Type C [Flavor$=QQQ'q\bar{q}$, J=1/2]:
\begin{eqnarray}\label{CMI3}
{{H}}_{CMI}^C=
\begin{matrix}
\left(
\begin{array}{ccccc}
2(2C_{12}+C_{45})&\frac{\sqrt3}{21}(15\gamma+13\delta)&2(\mu-\nu)&-\frac{\sqrt3}{21}(15\alpha-13\beta)&-\frac{\sqrt6}{21}(15\alpha-13\beta)\\
&\frac{2(7\lambda-13\gamma-15\delta)}{21}&-\frac{\sqrt3}{21}(15\alpha-13\beta)&\frac{42(\mu-\nu)-2(13\alpha-15\beta)}{21}&\frac{\sqrt2}{21}(13\alpha-15\beta)\\
&&\frac23(8\lambda-8\mu-3\nu)&\frac{2\sqrt3}{9}(2\beta-\delta)&-\frac{2\sqrt6}{9}(\beta-2\delta)\\
&&&\frac29(3\nu-4\alpha+2\gamma)&-\frac{2\sqrt2}{9}(\alpha-2\gamma)\\
&&&&\frac29(3\mu-5\alpha-5\gamma)
\end{array}\right),
\end{matrix}
\end{eqnarray}
where the base vector is ($\Phi_1^C$, $\Phi_2^C$, $\Phi_3^C$, $\Phi_4^C$, $\Phi_5^C$). The defined variables are $\alpha=7C_{14}+2C_{15}$, $\beta=7C_{14}-2C_{15}$, $\gamma=2C_{34}+7C_{35}$, $\delta=2C_{34}-7C_{35}$, $\mu=4C_{12}-C_{13}-C_{45}$, $\nu=4C_{12}+2C_{13}-C_{45}$, and $\lambda=6C_{12}-C_{45}$.
\item Type D [Flavor$=QQQ'q\bar{q}$, J=3/2]:
\begin{eqnarray}\label{CMI4}
{{H}}_{CMI}^D=
\begin{matrix}
\left(\begin{array}{cccc}
\frac{1}{21}(14\lambda+13\gamma+15\delta)&\frac{42(\mu-\nu)+(13\alpha-15\beta)}{21}&\frac{\sqrt3}{21}(15\alpha-13\beta)&\frac{\sqrt5}{21}(13\alpha-15\beta)\\
&\frac29(3\nu+2\alpha-\gamma)& \frac{2\sqrt3}{9}(\beta-2\delta) & -\frac{2\sqrt5}{9}(\alpha-2\gamma)\\
&& \frac23(8\lambda-4\nu-7\mu)& \frac{2\sqrt{15}}{9}(\beta+\delta)\\
&&&\frac29(3\mu-2\alpha-2\gamma)
\end{array}\right),
\end{matrix}
\end{eqnarray}
where the base vector is ($\Phi_1^D$, $\Phi_2^D$, $\Phi_3^D$, $\Phi_4^D$).
\item Type E [Flavor$=QQQ'q\bar{q}$, J=5/2]:
\begin{eqnarray}\label{CMI5}
{{H}}_{CMI}^E=
\frac{2}{3}(4C_{12}-C_{45}-C_{13}+7C_{14}+2C_{15}+2C_{34}+7C_{35}).
\end{eqnarray}
Here the only base is $\Phi_1^E$.
\end{itemize}

Comparing the present CMI matrices with those in Ref. \cite{Wu:2017weo}, one finds that the above expressions form a subset of those for the $qqqQ\bar{Q}$ case. The reason is that the flavor wave function of two identical heavy quarks must be symmetric while that of light quarks can also be antisymmetric. We will get the eigenvalues and eigenvectors of these matrices in the numerical evaluation. For the type-X pentaquarks, their mass splittings are the differences between the $N_X$ eigenvalues.

We use masses of conventional hadrons to determine relevant parameters. For convenience, here we also present CMI expressions for them \cite{Buccella:2006fn},
\begin{eqnarray}\label{b-m-CMI}
H_{CMI}(q_1\bar{q}_2)^{J=1}&=&\frac{16}{3}C_{12},\nonumber\\
H_{CMI}(q_1\bar{q}_2)^{J=0}&=&-16C_{12}, \nonumber\\
H_{CMI}(q_1q_2q_3)^{J=3/2}&=&\frac83(C_{12}+C_{23}+C_{13}), \nonumber\\
H_{CMI}(q_1q_2q_3)^{J=1/2}&=&\frac83\left[\begin{array}{cc}(C_{12}-2C_{23}-2C_{13})&\sqrt{3}(C_{23}-C_{13})\\\sqrt{3}(C_{23}-C_{13})&-3C_{12}\end{array}\right],
\end{eqnarray}
where the two bases for the last matrix correspond to the case of $J_{q_1q_2}=1$ and that of $J_{q_1q_2}=0$.

\section{Numerical analysis}\label{sec4}

\subsection{Parameter selection and estimation strategy}

\begin{table}[!htb]
\caption{Relevant coupling parameters in units of MeV.}\label{Cij}
\centering
\begin{tabular}{cc|cc}
\thickhline
$C_{cn}=4.0$ & $C_{cc}=5.3$ & $C_{c\bar{n}}=6.6$ & $C_{n\bar{n}}=29.8$\\
$C_{cs}=4.5$ & $C_{bc}=3.3$ & $C_{c\bar{s}}=6.7$ & $C_{n\bar{s}}=18.7$\\
$C_{bn}=1.3$ & $C_{bb}=2.9$ & $C_{b\bar{n}}=2.1$ & $C_{s\bar{s}}=6.5$\\
$C_{bs}=1.2$ &              & $C_{b\bar{s}}=2.3$\\
\thickhline
\end{tabular}
\end{table}

In order to estimate the masses of the possible pentaquark states, one needs to know the values of relevant mass parameters and coupling strengths. We extract the values of $C_{ij}$'s from the mass splittings of conventional hadrons. For example, $C_{nn}=1/16(m_{\Delta}-m_N)$ and $C_{n\bar{n}}=3/64(m_{\rho}-m_{\pi})$ may be determined with Eq. \eqref{b-m-CMI}. However, several effective coupling constants, $C_{s\bar{s}}$, $C_{cc}$, $C_{bb}$, $C_{bc}$, and $C_{b\bar{c}}$, need to be assigned by models or assumptions. Although the $\Xi_{cc}$ state has been observed, the extraction of $C_{cc}$ needs more measured baryon masses. We here simply adopt the assumption $C_{cc}$=$C_{c\bar{c}}$ for our evaluations. Similarly, we use $C_{s\bar{s}}$=$C_{ss}$, $C_{bb}$=$C_{b\bar{b}}$, and $C_{bc}$=$C_{b\bar{c}}$, where $C_{b\bar{c}}$ is obtained with the mass difference between $B_c^*$ and $B_c$ \cite{Godfrey:1985xj}. A variation of the values related with heavy quarks does not induce significant differences \cite{Wu:2016vtq,Lee:2007tn,Lee:2009rt}. Table \ref{Cij} lists all the coupling strengths we will use. At present, we further assume that these parameters can be applied to different systems so that we may estimate the masses of the studied pentaquark states. In the last part of this section, we will check the effects caused by the adoption of other values of coupling parameters.

According to Eq. \eqref{CMI}, the mass of a pentquark state in the chromomagnetic model is
\begin{eqnarray}\label{mass}
M=\sum_{i=1}^5 m_i+\langle \widehat{H}_{CMI}\rangle,
\end{eqnarray}
where $M$ and $\langle\widehat{H}_{CMI}\rangle$ are the mass of a pentaquark and the corresponding eigenvalue of the choromomagnetic interaction, respectively. By introducing a reference system, the mass of the pentaquark state can be written as
\begin{eqnarray}\label{massref}
 M=(M_{ref}-\langle\widehat{H}_{CMI}\rangle_{ref})+\langle\widehat{H}_{CMI}\rangle.
\end{eqnarray}
Here, $M_{ref}$ and $\langle\widehat{H}_{CMI}\rangle_{ref}$ are the mass of the reference system and the corresponding chromomagnetic interaction, respectively. Because none of $QQQq\bar{q}$ pentaquark states is observed, we choose the threshold of a baryon-meson state with the same quark content as $M_{ref}$. If the simple model could give correct masses for all the hadron states, the above two formulas should be equivalent. In fact, the model does not involve dynamics and the two approaches result in different multiquark masses. One may consult Refs. \cite{Wu:2017weo,Zhou:2018pcv} for some discussions on the difference.

\begin{table}[!htb]
\caption{Used masses of the conventional hadrons in units of MeV \cite{Tanabashi:2018oca}. Since the spin of the $\Xi_{cc}$ observed by LHCb may be 1/2 or 3/2, we show results in both cases. The adopted masses of other doubly heavy baryons are taken from Ref. \cite{Gershtein:1998un} ($\Omega_{cc}^*$ is from Ref. \cite{Weng:2018mmf}). The values in parentheses are obtained with the parameters in table \ref{Cij}.}\label{Hadron}
\centering
\begin{tabular}{p{0.80cm}<{\centering}p{1.70cm}<{\centering}"p{0.80cm}<{\centering}p{1.70cm}<{\centering}"p{0.80cm}<{\centering}p{1.70cm}<{\centering}"p{0.80cm}<{\centering}p{1.70cm}<{\centering}}
\thickhline
\multicolumn{2}{c"}{Mesons (J=0)}&\multicolumn{2}{c"}{Mesons (J=1)}&\multicolumn{2}{c"}{Baryons (J=1/2)}&\multicolumn{2}{c}{Baryons (J=3/2)}\\
$\pi$   & 139.6&$\rho  $  & 775.3 &$N$              & 938.3   &$\Delta$       &1232.0   \\
        &      &$\omega$  & 782.7 &$\Xi$            &1314.9   &$\Xi^*$        &1531.8   \\
        &      &$\phi  $  &1019.5 &                 &         &$\Omega$       &1672.5   \\
$K$     & 493.7&$K^{*}$   & 891.8 &$\Sigma_c$       &2454.0   &$\Sigma_c^*$   &2518.4   \\
$D  $   &1869.7&$D^*  $   &2010.3 &$\Xi_c^\prime$   &2577.4   &$\Xi_c^*$      &2645.5   \\
$D_s$   &1968.3&$D_s^*$   &2112.2 &$\Sigma_b$       &5811.3   &$\Sigma_b^*$   &5832.1   \\
$B  $   &5279.3&$B^*  $   &5324.7 &$\Xi_b^\prime$   &5935.0   &$\Xi_b^*$      &5955.3   \\
$B_s$   &5366.9&$B_s^*$   &5415.4 &$\Xi_{cc}$       &3621.4   &$\Xi_{cc}^*$   &(3685.4) \\
$\eta_c$&2983.9&$J/\psi$  &3096.9 &$\Xi_{cc}$       &(3557.4) &$\Xi_{cc}^*$   &3621.4   \\
$\eta_b$&9399.0&$\Upsilon$&9460.3 &$\Omega_{cc}$    &(3730.4) &$\Omega_{cc}^*$&3802.4   \\
        &      &$        $&       &$\Xi_{bb}   $    &10093.0  &$\Xi_{bb}^*   $&(10113.8)\\
$   $   &      &$        $&       &$\Omega_{bb}$    &10193.0  &$\Omega_{bb}^*$&(10212.2)\\
$   $   &      &$        $&       &$\Xi_{bc}$       &6820.0                             \\
$   $   &      &$        $&       &$\Xi_{bc}'$      &(6845.9) &$\Xi_{bc}^*$   &(6878.8) \\
$   $   &      &$        $&       &$\Omega_{bc}$    &6920.0                             \\
$   $   &      &$        $&       &$\Omega_{bc}'$   &(6950.9) &$\Omega_{bc}^*$&(6983.4) \\
\thickhline
\end{tabular}
\end{table}

\begin{table}[!htb]
\caption{The masses of triply heavy baryons in units of MeV in the literature. One may consult Refs. \cite{Wei:2015gsa,Wei:2016jyk} for more results about the $\Omega_{ccc}$ and $\Omega_{bbb}$.}\label{QQQ-sum}
\centering
\begin{tabular}{p{1.6cm}<{\centering}p{2.3cm}<{\centering}p{2.3cm}<{\centering}p{2.3cm}<{\centering}p{2.3cm}<{\centering}p{2.3cm}<{\centering}p{2.3cm}<{\centering}}
\thickhline
\hline
Baryon                          &$\Omega_{ccc}$ &$\Omega^*_{ccb}$&$\Omega_{ccb}$&$\Omega^*_{bbc}$&$\Omega_{bbc}$ &$\Omega_{bbb}$\\\hline
Ref. \cite{Ponce:1978gk}        &$-$            &8039           &$-$            &11152           &$-$             &14248           \\
Ref. \cite{Hasenfratz:1980ka}   &4790           &8030           &$-$            &11200           &$-$             &14300           \\
Ref. \cite{Bjorken:1985ei}      &$4925\pm90$    &$8200\pm90$    &$-$            &$11480\pm120$   &$-$             &$14760\pm180$   \\
Ref. \cite{Tsuge:1985ei}        &4808           &7852           &7828           &10854           &10827           &13823           \\
Ref. \cite{SilvestreBrac:1996bg}&4796$\sim$4806 &$-$            &$7984\sim8032$ &$-$             &$11163\sim11220$&$14348\sim14398$\\
Ref. \cite{Itoh:2000um}         &4847.6         &$-$            &$-$            &$-$             &$-$             &$-$             \\
Ref. \cite{Faessler:2006ft}     &$-$            &$-$            &8000           &$-$             &11500           &$-$             \\
Ref. \cite{Jia:2006gw}          &$4760\pm60$    &$7980\pm70$    &$7980\pm70$    &$11190\pm80$    &$11190\pm80$    &$14370\pm80$    \\
Ref. \cite{Roberts:2007ni}      &4965           &8265           &8245           &11554           &11535           &14834           \\
Ref. \cite{Bernotas:2008bu}     &4777           &8005           &7984           &11163           &11139           &14276           \\
Ref. \cite{Martynenko:2007je}   &4803           &8025           &8018           &11287           &11280           &14569           \\
Ref. \cite{Zhang:2009re}        &$4670\pm150$   &$7450\pm160$   &$7410\pm130$   &$10540\pm110$   &$10300\pm100$   &$13280\pm100$   \\
Ref. \cite{Meinel:2010pw}       &$-$            &$-$            &$-$            &$-$             &$-$             &$14371\pm12$    \\
Ref. \cite{Wang:2011ae}         &$4990\pm140$   & $8230\pm130$  &$8230\pm130$   &$11490\pm110$   &$11500\pm110$   &$14830\pm100$   \\
Ref. \cite{Durr:2012dw}         &$4774\pm24$    &$-$            &$-$            &$-$             &$-$             &$-$             \\
Ref. \cite{Flynn:2011gf}        &4799           &8046           &8018           &11245           &11214           &14398           \\
Ref. \cite{Meinel:2012qz}     &$4990\pm140$   &$8230\pm130$   &$8230\pm130$   &$11490\pm110$   &$11500\pm110$   &$14830\pm100$   \\
Ref. \cite{Aliev:2012tt}        &$-$            &$-$            &$8500\pm120$   &$-$             &$11730\pm160$\\
Ref. \cite{Aliev:2014lxa}       &$4720\pm120$   &$8070\pm100$   &$-$            &$11350\pm150$   &$-$             &$14300\pm200$   \\
Ref. \cite{Namekawa:2013vu}     &$4789\pm6\pm21$&$-$            &$-$            &$-$             &$-$             &$-$             \\
Ref. \cite{Dhir:2013nka}        &$-$            &8465           &8463           &11797           &11795           &15129           \\
Ref. \cite{Brown:2014ena}       &$4796\pm8\pm18$&$8037\pm9\pm20$&$8007\pm9\pm20$&$11229\pm8\pm20$&$11195\pm8\pm20$&$14366\pm9\pm20$\\
Ref. \cite{Padmanath:2015jea}   &$-$            &$8050\pm10$    &$-$            &$-$             &$-$             &$-$             \\
Ref. \cite{Can:2015exa}         &$4769\pm6$     &$-$            &$-$            &$-$             &$-$             &$-$             \\
Ref. \cite{Thakkar:2016sog}     &4760           &$8027\sim8032$ &$7999\sim8005$ &$11284\sim11287$&$11274\sim11277$&14370           \\
Ref. \cite{Shah:2017jkr}        &4806           &$-$            &$-$            &$-$             &$-$             &14496           \\
Ref. \cite{Weng:2018mmf}        &4785.6         &8021.8         &7990.3         &11196.4         &11165.0         &14309.7         \\
Ref. \cite{Mathur:2018epb}      &$-$            &$8026\pm7\pm11$&$8005\pm6\pm11$&$11211\pm6\pm12$&$11194\pm5\pm12$&$-$             \\
Ref. \cite{Shah:2018div}     &$-$            &$-$            &$-$            &11296           &11231           &$-$             \\
\thickhline
\end{tabular}
\end{table}

If we adopt Eq. \eqref{mass}, one needs to know the effective quark masses. Their values extracted from lowest conventional hadrons are $m_n = 361.7$ MeV, $m_s = 540.3$ MeV, $m_c = 1724.6$ MeV, and $m_b = 5052.8$ MeV. From Eq. \eqref{mass} and the coupling parameters in table \ref{Cij}, the masses of conventional hadrons may be evaluated. By comparing them with the experimental measurements, one finds overestimated theoretical results \cite{Wu:2016gas,Zhou:2018pcv}. The multiquark masses are also generally overestimated \cite{Wu:2017weo,Wu:2016vtq,Chen:2016ont,Luo:2017eub,Zhou:2018pcv}. Thus we may treat the masses obtained in this method as theoretical upper limits. This fact indicates that attractions inside hadrons cannot be taken into account sufficiently in the simple chromomagnetic model. A more reasonable approach is to adopt Eq. \eqref{massref}, where the necessary attractions for conventional hadrons have been implicitly included in their physical masses. In this approach, probably there are two choices on the reference threshold. In this case, we present results with both thresholds. To calculate the baryon-meson thresholds, we use the hadron masses shown in table \ref{Hadron}. The extraction of the above effective quark masses and coupling parameters also relies on these numbers. Here, we assume that the spin of $\Xi_{cc}$ is 1/2, which is presumed in Ref. \cite{Chen:2017sbg}. If the spin of $\Xi_{cc}$ is 3/2 (consistent with the prediction in Ref. \cite{Gershtein:1998un}), the multiquark masses estimated with the threshold relating to $\Xi_{cc}$ would be shifted downward by 64 MeV. When determining thresholds relating to other doubly heavy baryons, we adopt the values obtained in Ref. \cite{Gershtein:1998un}. However, the mass of $\Omega_{cc}^*$ in that reference was obtained by adding $\sim$100 MeV to the mass of $\Xi^*_{cc}$. Consider the fact that the quantum numbers of the observed $\Xi_{cc}$ by LHCb have not been determined, we prefer to use another value obtained in Ref. \cite{Weng:2018mmf}. A different mass of the doubly heavy baryon (see Refs. \cite{Wei:2015gsa,Wei:2016jyk} for a collection) will lead to a different mass of the pentaquark. One may compare the adopted mass with the value in table \ref{Hadron} to estimate the mass difference. For example, if we got a pentaquark mass $M_1$ with $M_{\Omega_{bc}}=6920$ MeV while one wants to adopt $M_{\Omega_{bc}^*}=7065.7$ MeV \cite{Weng:2018mmf}, then one would obtain the mass of that pentaquark by adding $(7065.7-6983.4)$ MeV to $M_1$, where 6983.4 MeV is the mass of $\Omega_{bc}^*$ in table \ref{Hadron}. In order for further discussions, we also present a summary for the theoretical investigations in the literature on the masses of triply heavy conventional baryons in Table \ref{QQQ-sum}.

\subsection{Numerical results and global features}

Based on the content of heavy quarks, one may classify the pentaquarks into four groups: $cccq\bar{q}$, $bbbq\bar{q}$, $ccbq\bar{q}$, and $bbcq\bar{q}$ or two types: $QQQq\bar{q}$ and $QQQ'q\bar{q}$. With all the parameters given above, we get the spectra of these pentaquark systems which are shown in tables \ref{penta-massccc}-\ref{penta-massbbc}. In these tables, the columns labeled with the title ``Mass'' list masses estimated with Eq. \eqref{mass}. The last one or two columns show results estimated with relevant baryon-meson thresholds. For the $QQQ'q\bar{q}$ states, one may use $(QQq)$-$(Q'\bar{q})$ or $(QQ'q)$-$(Q\bar{q})$ type threshold for our purpose. The resulting masses may have an uncertainty about 100 MeV, which is not a very large value if compared to the mass around 10 GeV. We present results with both type thresholds.

\begin{table}[htbp]
\center\caption{Numerical results for the $cccq\bar{q}$ systems in units of MeV. The masses in the sixth column are estimated with Eq. \eqref{mass} and those in the last column with the $(ccq)$-$(c\bar{q})$ type threshold.}\label{penta-massccc}
\begin{tabular}{ccccccc}\hline
System & $J^P$ & $\langle H_{CM} \rangle$ & Eigenvalues & Eigenvectors & Mass &$\Xi_{cc}D$ \\
$cccn\bar{n}$& $\frac12^-$ &$\left[\begin{array}{cc}112.6&61.2\\61.2&50.5\end{array}\right]$&$\left[\begin{array}{c}150.2\\12.9\end{array}\right]$&$\left[\begin{array}{c}(0.85,0.52)\\(0.52,-0.85)\end{array}\right]$&$\left[\begin{array}{c}6047\\5910\end{array}\right]$&$\left[\begin{array}{c}5775\\5638\end{array}\right]$\\
&$\frac32^-$ &24.5&24.5&1&5922&5650\\ \hline
System & $J^P$ & $\langle H_{CM} \rangle$ & Eigenvalues & Eigenvectors & Mass &$\Xi_{cc}D_s$ \\
$cccn\bar{s}$& $\frac12^-$ &$\left[\begin{array}{cc}90.4&61.8\\61.8&58.5\end{array}\right]$&$\left[\begin{array}{c}138.3\\10.7\end{array}\right]$&$\left[\begin{array}{c}(0.79,0.61)\\(0.61,-0.79)\end{array}\right]$&$\left[\begin{array}{c}6214\\6086\end{array}\right]$&$\left[\begin{array}{c}5864\\5736\end{array}\right]$\\
&$\frac32^-$ &31.5&31.5&1&6107&5757\\ \hline
System & $J^P$ & $\langle H_{CM} \rangle$ & Eigenvalues & Eigenvectors & Mass &$\Omega_{cc}D$ \\
$cccs\bar{n}$& $\frac12^-$ &$\left[\begin{array}{cc}90.4&64.1\\64.1&54.5\end{array}\right]$&$\left[\begin{array}{c}139.0\\5.9\end{array}\right]$&$\left[\begin{array}{c}(0.80,0.60)\\(0.60,-0.80)\end{array}\right]$&$\left[\begin{array}{c}6215\\6082\end{array}\right]$&$\left[\begin{array}{c}5879\\5745\end{array}\right]$\\
&$\frac32^-$ &33.5&33.5&1&6109&5773\\ \hline
System & $J^P$ & $\langle H_{CM} \rangle$ & Eigenvalues & Eigenvectors & Mass &$\Omega_{cc}D_s$ \\
$cccs\bar{s}$& $\frac12^-$ &$\left[\begin{array}{cc}66.0&64.7\\64.7&63.3\end{array}\right]$&$\left[\begin{array}{c}129.3\\-0.0\end{array}\right]$&$\left[\begin{array}{c}(0.71,0.70)\\(0.70,-0.71)\end{array}\right]$&$\left[\begin{array}{c}6384\\6254\end{array}\right]$&$\left[\begin{array}{c}5969\\5840\end{array}\right]$\\
&$\frac32^-$ &41.3&41.3&1&6296&5881\\ \hline
\end{tabular}
\end{table}

\begin{table}[htbp]
\center\caption{Numerical results for the $bbbq\bar{q}$ systems in units of MeV. The masses in the sixth column are estimated with Eq. \eqref{mass} and those in the last column with the $(bbq)$-$(b\bar{q})$ type threshold.}\label{penta-massbbb}
\begin{tabular}{ccccccc}\hline
System & $J^P$ & $\langle H_{CM} \rangle$ & Eigenvalues & Eigenvectors & Mass &$\Xi_{bb}\bar{B}$ \\
$bbbn\bar{n}$& $\frac12^-$ &$\left[\begin{array}{cc}88.6&19.6\\19.6&14.5\end{array}\right]$&$\left[\begin{array}{c}93.5\\9.6\end{array}\right]$&$\left[\begin{array}{c}(0.97,0.24)\\(0.24,-0.97)\end{array}\right]$&$\left[\begin{array}{c}15975\\15891\end{array}\right]$&$\left[\begin{array}{c}15506\\15422\end{array}\right]$\\
&$\frac32^-$ &6.5&6.5&1&15888&15418\\ \hline
System & $J^P$ & $\langle H_{CM} \rangle$ & Eigenvalues & Eigenvectors & Mass &$\Xi_{bb}\bar{B}_s$ \\
$bbbn\bar{s}$& $\frac12^-$ &$\left[\begin{array}{cc}66.4&20.8\\20.8&23.2\end{array}\right]$&$\left[\begin{array}{c}74.8\\14.8\end{array}\right]$&$\left[\begin{array}{c}(0.93,0.37)\\(0.37,-0.93)\end{array}\right]$&$\left[\begin{array}{c}16135\\16075\end{array}\right]$&$\left[\begin{array}{c}15578\\15518\end{array}\right]$\\
&$\frac32^-$ &13.2&13.2&1&16074&15516\\ \hline
System & $J^P$ & $\langle H_{CM} \rangle$ & Eigenvalues & Eigenvectors & Mass &$\Omega_{bb}\bar{B}$ \\
$bbbs\bar{n}$& $\frac12^-$ &$\left[\begin{array}{cc}66.4&19.1\\19.1&22.5\end{array}\right]$&$\left[\begin{array}{c}73.5\\15.4\end{array}\right]$&$\left[\begin{array}{c}(0.94,0.35)\\(0.35,-0.94)\end{array}\right]$&$\left[\begin{array}{c}16134\\16076\end{array}\right]$&$\left[\begin{array}{c}15584\\15526\end{array}\right]$\\
&$\frac32^-$ &13.5&13.5&1&16074&15524\\ \hline
System & $J^P$ & $\langle H_{CM} \rangle$ & Eigenvalues & Eigenvectors & Mass &$\Omega_{bb}\bar{B}_s$ \\
$bbbs\bar{s}$& $\frac12^-$ &$\left[\begin{array}{cc}42.0&20.2\\20.2&32.0\end{array}\right]$&$\left[\begin{array}{c}57.8\\16.2\end{array}\right]$&$\left[\begin{array}{c}(0.79,0.62)\\(0.62,-0.79)\end{array}\right]$&$\left[\begin{array}{c}16297\\16255\end{array}\right]$&$\left[\begin{array}{c}15660\\15618\end{array}\right]$\\
&$\frac32^-$ &21.0&21.0&1&16260&15623\\ \hline
\end{tabular}
\end{table}

\begin{table}[htbp]
\center\caption{Numerical results for the $ccbq\bar{q}$ systems in units of MeV. The masses in the sixth column are estimated with Eq. \eqref{mass} and those in the last two columns with the $(ccq)$-$(b\bar{q})$ type and $(bcq)$-$(c\bar{q})$ type thresholds, respectively.}\label{penta-massccb}\footnotesize
\begin{tabular}{cccccccc}\hline
System & $J^P$ & $\langle H_{CM} \rangle$ & Eigenvalues & Eigenvectors & Mass &$\Xi_{cc}\bar{B}$&$\Xi_{bc}D$ \\
$ccbn\bar{n}$& $\frac12^-$ &$\left[\begin{array}{ccccc}80.8&8.4&-19.8&-35.1&-49.6\\8.4&-2.8&-35.1&-49.7&21.1\\-19.8&-35.1&78.1&16.1&-21.2\\-35.1&-49.7&16.1&-30.3&-2.1\\-49.6&21.1&-21.2&-2.1&-72.9\end{array}\right]$&$\left[\begin{array}{c}124.6\\83.6\\12.1\\-72.3\\-95.0\end{array}\right]$&$\left[\begin{array}{c}(-0.66,-0.32,0.60,0.31,0.06)\\(0.66,-0.27,0.61,0.00,-0.33)\\(-0.12,0.68,0.49,-0.52,0.13)\\(0.18,0.57,0.08,0.80,0.05)\\(0.29,-0.19,0.11,0.01,0.93)\end{array}\right]$&$\left[\begin{array}{c}9350\\9309\\9238\\9153\\9130\end{array}\right]$&$\left[\begin{array}{c}9087\\9046\\8975\\8891\\8868\end{array}\right]$&$\left[\begin{array}{c}8956\\8915\\8843\\8759\\8736\end{array}\right]$\\
& $\frac32^-$ &$\left[\begin{array}{cccc}3.4&-4.9&35.1&33.4\\-4.9&13.1&15.0&-3.3\\35.1&15.0&71.5&2.3\\33.4&-3.3&2.3&-33.9\end{array}\right]$&$\left[\begin{array}{c}90.0\\21.6\\-1.7\\-55.7\end{array}\right]$&$\left[\begin{array}{c}(0.40,0.14,0.90,0.12)\\(0.53,-0.75,-0.17,0.36)\\(0.50,0.65,-0.39,0.42)\\(-0.55,-0.03,0.14,0.82)\end{array}\right]$&$\left[\begin{array}{c}9315\\9247\\9224\\9170\end{array}\right]$&$\left[\begin{array}{c}9053\\8984\\8961\\8907\end{array}\right]$&$\left[\begin{array}{c}8921\\8853\\8829\\8775\end{array}\right]$\\
& $\frac52^-$ &31.1&31.1&1&9256&8994&8862\\ \hline
System & $J^P$ & $\langle H_{CM} \rangle$ & Eigenvalues & Eigenvectors & Mass &$\Xi_{cc}\bar{B}_s$&$\Xi_{bc}D_s$ \\
$ccbn\bar{s}$& $\frac12^-$ &$\left[\begin{array}{ccccc}58.6&8.7&-19.8&-35.6&-50.3\\8.7&4.9&-35.6&-50.2&21.5\\-19.8&-35.6&55.9&16.4&-22.6\\-35.6&-50.2&16.4&-22.4&-1.3\\-50.3&21.5&-22.6&-1.3&-67.3\end{array}\right]$&$\left[\begin{array}{c}110.9\\68.6\\9.7\\-66.3\\-93.3\end{array}\right]$&$\left[\begin{array}{c}(-0.59,-0.42,0.58,0.38,0.04)\\(-0.70,0.32,-0.51,0.00,0.39)\\(-0.13,0.60,0.62,-0.48,0.08)\\(0.19,0.58,0.09,0.79,-0.02)\\(0.34,-0.16,0.14,0.04,0.92)\end{array}\right]$&$\left[\begin{array}{c}9515\\9473\\9414\\9338\\9311\end{array}\right]$&$\left[\begin{array}{c}9165\\9122\\9063\\8987\\8960\end{array}\right]$&$\left[\begin{array}{c}9042\\9000\\8941\\8865\\8838\end{array}\right]$\\
& $\frac32^-$ &$\left[\begin{array}{cccc}10.7&-4.6&35.6&34.0\\-4.6&20.3&16.0&-2.0\\35.6&16.0&49.3&0.9\\34.0&-2.0&0.9&-27.2\end{array}\right]$&$\left[\begin{array}{c}76.0\\27.5\\0.4\\-50.8\end{array}\right]$&$\left[\begin{array}{c}(0.52,0.18,0.81,0.18)\\(0.47,-0.80,-0.19,0.32)\\(0.44,0.56,-0.51,0.48)\\(-0.56,-0.06,0.20,0.80)\end{array}\right]$&$\left[\begin{array}{c}9480\\9431\\9404\\9353\end{array}\right]$&$\left[\begin{array}{c}9130\\9081\\9054\\9003\end{array}\right]$&$\left[\begin{array}{c}9007\\8959\\8932\\8881\end{array}\right]$\\
&$\frac52^-$ &39.5&39.5&1&9444&9093&8971\\ \hline
System & $J^P$ & $\langle H_{CM} \rangle$ & Eigenvalues & Eigenvectors & Mass &$\Omega_{cc}\bar{B}$&$\Omega_{bc}D$ \\
$ccbs\bar{n}$& $\frac12^-$ &$\left[\begin{array}{ccccc}58.6&8.0&-19.8&-35.7&-50.5\\8.0&5.1&-35.7&-49.0&20.6\\-19.8&-35.7&55.9&18.8&-23.4\\-35.7&-49.0&18.8&-26.1&-3.3\\-50.5&20.6&-23.4&-3.3&-69.2\end{array}\right]$&$\left[\begin{array}{c}110.8\\68.9\\6.9\\-67.1\\-95.1\end{array}\right]$&$\left[\begin{array}{c}(-0.58,-0.41,0.59,0.38,0.03)\\(0.71,-0.32,0.49,0.01,-0.39)\\(-0.11,0.62,0.62,-0.46,0.07)\\(0.16,0.57,0.05,0.80,-0.09)\\(0.35,-0.11,0.15,0.10,0.91)\end{array}\right]$&$\left[\begin{array}{c}9515\\9473\\9411\\9337\\9309\end{array}\right]$&$\left[\begin{array}{c}9188\\9146\\9084\\9010\\8982\end{array}\right]$&$\left[\begin{array}{c}9046\\9004\\8942\\8868\\8840\end{array}\right]$\\
& $\frac32^-$ &$\left[\begin{array}{cccc}10.5&-5.2&35.7&32.6\\-5.2&22.1&16.5&-5.2\\35.7&16.5&49.3&5.2\\32.6&-5.2&5.2&-28.0\end{array}\right]$&$\left[\begin{array}{c}76.9\\30.0\\-4.1\\-48.7\end{array}\right]$&$\left[\begin{array}{c}(0.52,0.18,0.81,0.19)\\(-0.43,0.83,0.16,-0.30)\\(0.49,0.52,-0.53,0.45)\\(-0.55,-0.02,0.16,0.82)\end{array}\right]$&$\left[\begin{array}{c}9481\\9434\\9400\\9355\end{array}\right]$&$\left[\begin{array}{c}9154\\9107\\9073\\9028\end{array}\right]$&$\left[\begin{array}{c}9012\\8965\\8931\\8886\end{array}\right]$\\
&$\frac52^-$ &40.7&40.7&1&9445&9118&8975\\ \hline
System & $J^P$ & $\langle H_{CM} \rangle$ & Eigenvalues & Eigenvectors & Mass &$\Omega_{cc}\bar{B}_s$&$\Omega_{bc}D_s$ \\
$ccbs\bar{s}$& $\frac12^-$ &$\left[\begin{array}{ccccc}34.2&8.2&-19.8&-36.1&-51.1\\8.2&13.5&-36.1&-49.5&21.0\\-19.8&-36.1&31.5&19.2&-24.8\\-36.1&-49.5&19.2&-17.5&-2.5\\-51.1&21.0&-24.8&-2.5&-62.8\end{array}\right]$&$\left[\begin{array}{c}100.5\\54.6\\-0.9\\-60.4\\-94.9\end{array}\right]$&$\left[\begin{array}{c}(0.47,0.53,-0.54,-0.45,0.01)\\(-0.75,0.34,-0.34,0.04,0.46)\\(0.13,-0.52,-0.74,0.41,0.00)\\(0.18,0.58,0.06,0.78,-0.15)\\(0.42,-0.08,0.20,0.12,0.88)\end{array}\right]$&$\left[\begin{array}{c}9683\\9637\\9582\\9522\\9488\end{array}\right]$&$\left[\begin{array}{c}9268\\9223\\9167\\9108\\9073\end{array}\right]$&$\left[\begin{array}{c}9135\\9090\\9034\\8975\\8940\end{array}\right]$\\
& $\frac32^-$ &$\left[\begin{array}{cccc}18.5&-4.9&36.1&33.2\\-4.9&30.0&17.5&-3.9\\36.1&17.5&24.9&3.8\\33.2&-3.9&3.8&-20.6\end{array}\right]$&$\left[\begin{array}{c}67.3\\36.6\\-7.2\\-43.8\end{array}\right]$&$\left[\begin{array}{c}(0.66,0.20,0.67,0.27)\\(-0.35,0.89,0.18,-0.25)\\(0.34,0.41,-0.66,0.52)\\(-0.57,-0.06,0.28,0.77)\end{array}\right]$&$\left[\begin{array}{c}9650\\9619\\9575\\9539\end{array}\right]$&$\left[\begin{array}{c}9235\\9205\\9161\\9124\end{array}\right]$&$\left[\begin{array}{c}9102\\9072\\9028\\8991\end{array}\right]$\\
&$\frac52^-$ &49.9&49.9&1&9632&9218&9085\\ \hline
\end{tabular}
\end{table}

\begin{table}[htbp]
\center\caption{Numerical results for the $bbcq\bar{q}$ systems in units of MeV. The masses in the sixth column are estimated with Eq. \eqref{mass} and those in the last two columns with the $(bcq)$-$(b\bar{q})$ type and $(bbq)$-$(c\bar{q})$ type thresholds, respectively.}\label{penta-massbbc}\footnotesize
\begin{tabular}{cccccccc}\hline
System & $J^P$&$\langle H_{CM}\rangle$&Eigenvalues&Eigenvectors&Mass&$\Xi_{bc}\bar{B}$&$\Xi_{bb}D$\\
$bbcn\bar{n}$& $\frac12^-$&$\left[\begin{array}{ccccc}71.2&26.1&-19.8&-11.2&-15.8\\26.1&-20.8&-11.2&-29.3&6.7\\-19.8&-11.2&71.7&18.5&-44.3\\-11.2&-29.3&18.5&4.5&29.9\\-15.8&6.7&-44.3&29.9&-89.3\end{array}\right]$&$\left[\begin{array}{c}105.6\\65.0\\16.8\\-35.9\\-114.2\end{array}\right]$&$\left[\begin{array}{c}(0.63,0.25,-0.69,-0.25,0.08)\\(0.66,0.13,0.67,-0.12,-0.28)\\(0.37,-0.34,-0.07,0.84,0.19)\\(-0.12,0.88,0.11,0.36,0.26)\\(0.11,-0.16,0.25,-0.29,0.90)\end{array}\right]$&$\left[\begin{array}{c}12659\\12619\\12570\\12518\\12439\end{array}\right]$&$\left[\begin{array}{c}12274\\12234\\12186\\12133\\12055\end{array}\right]$&$\left[\begin{array}{c}12180\\12139\\12091\\12039\\11960\end{array}\right]$\\
& $\frac32^-$ &$\left[\begin{array}{cccc}-2.0&-15.1&11.2&10.6\\-15.1&-13.9&31.3&47.3\\11.2&31.3&65.1&-28.7\\10.6&47.3&-28.7&-44.3\end{array}\right]$&$\left[\begin{array}{c}77.8\\21.4\\-0.1\\-94.2\end{array}\right]$&$\left[\begin{array}{c}(0.07,0.25,0.96,-0.12)\\(0.30,-0.77,0.11,-0.55)\\(0.93,0.10,-0.05,0.36)\\(-0.21,-0.58,0.26,0.74)\end{array}\right]$&$\left[\begin{array}{c}12631\\12575\\12554\\12459\end{array}\right]$&$\left[\begin{array}{c}12247\\12190\\12169\\12075\end{array}\right]$&$\left[\begin{array}{c}12152\\12096\\12074\\11980\end{array}\right]$\\
& $\frac52^-$ &30.7&30.7&1&12584&12199&12105\\\hline
System & $J^P$&$\langle H_{CM}\rangle$&Eigenvalues&Eigenvectors&Mass&$\Xi_{bc}\bar{B}_s$&$\Xi_{bb}D_s$\\
$bbcn\bar{s}$& $\frac12^-$&$\left[\begin{array}{ccccc}49.0&26.2&-19.8&-12.1&-17.1\\26.2&-13.3&-12.1&-30.3&7.4\\-19.8&-12.1&49.5&18.4&-44.8\\-12.1&-30.3&18.4&11.9&30.2\\-17.1&7.4&-44.8&30.2&-83.2\end{array}\right]$&$\left[\begin{array}{c}89.5\\48.1\\17.4\\-29.9\\-111.2\end{array}\right]$&$\left[\begin{array}{c}(0.61,0.34,-0.63,-0.35,0.06)\\(0.53,0.17,0.71,-0.25,-0.36)\\(0.56,-0.25,-0.02,0.78,0.13)\\(-0.15,0.88,0.15,0.35,0.24)\\(0.13,-0.16,0.29,-0.29,0.89)\end{array}\right]$&$\left[\begin{array}{c}12822\\12780\\12750\\12702\\12621\end{array}\right]$&$\left[\begin{array}{c}12349\\12308\\12277\\12230\\12148\end{array}\right]$&$\left[\begin{array}{c}12264\\12223\\12192\\12145\\12063\end{array}\right]$\\
& $\frac32^-$ &$\left[\begin{array}{cccc}5.3&-14.5&12.1&11.8\\-14.5&-6.4&31.7&47.8\\12.1&31.7&42.9&-29.6\\11.8&47.8&-29.6&-37.4\end{array}\right]$&$\left[\begin{array}{c}59.4\\28.0\\7.9\\-90.9\end{array}\right]$&$\left[\begin{array}{c}(0.09,0.35,0.93,-0.10)\\(0.28,-0.74,0.20,-0.58)\\(0.93,0.06,-0.08,0.35)\\(0.21,0.57,-0.32,-0.73)\end{array}\right]$&$\left[\begin{array}{c}12792\\12760\\12740\\12641\end{array}\right]$&$\left[\begin{array}{c}12319\\12288\\12267\\12169\end{array}\right]$&$\left[\begin{array}{c}12234\\12203\\12183\\12084\end{array}\right]$\\
&$\frac52^-$ &38.8&38.8&1&12771&12298&12213\\\hline
System&$J^P$&$\langle H_{CM}\rangle$&Eigenvalues&Eigenvectors&Mass&$\Omega_{bc}\bar{B}$&$\Omega_{bb}D$\\
$bbcs\bar{n}$& $\frac12^-$ &$\left[\begin{array}{ccccc}49.0&28.4&-19.8&-11.1&-15.7\\28.4&-16.1&-11.1&-29.4&6.8\\-19.8&-11.1&49.5&17.6&-42.8\\-11.1&-29.4&17.6&13.0&30.7\\-15.7&6.8&-42.8&30.7&-82.3\end{array}\right]$&$\left[\begin{array}{c}88.7\\47.5\\18.3\\-32.1\\-109.2\end{array}\right]$&$\left[\begin{array}{c}(0.62,0.33,-0.62,-0.34,0.05)\\(0.50,0.18,0.72,-0.26,-0.35)\\(0.55,-0.20,0.02,0.80,0.14)\\(-0.19,0.89,0.14,0.31,0.25)\\(0.13,-0.16,0.28,-0.29,0.89)\end{array}\right]$&$\left[\begin{array}{c}12821\\12780\\12751\\12700\\12623\end{array}\right]$&$\left[\begin{array}{c}12361\\12320\\12291\\12240\\12163\end{array}\right]$&$\left[\begin{array}{c}12262\\12221\\12192\\12141\\12064\end{array}\right]$\\
& $\frac32^-$ &$\left[\begin{array}{cccc}6.7&-15.0&11.1&10.7\\-15.0&-7.0&30.3&48.6\\11.1&30.3&42.9&-28.4\\10.7&48.6&-28.4&-37.1\end{array}\right]$&$\left[\begin{array}{c}58.0\\29.2\\8.5\\-90.1\end{array}\right]$&$\left[\begin{array}{c}(0.08,0.35,0.93,-0.09)\\(-0.31,0.74,-0.19,0.57)\\(0.92,0.09,-0.08,0.36)\\(0.20,0.58,-0.30,-0.73)\end{array}\right]$&$\left[\begin{array}{c}12790\\12761\\12741\\12642\end{array}\right]$&$\left[\begin{array}{c}12330\\12302\\12281\\12182\end{array}\right]$&$\left[\begin{array}{c}12231\\12203\\12182\\12083\end{array}\right]$\\
&$\frac52^-$ &38.3&38.3&1&12770&12311&12212\\\hline
System&$J^P$&$\langle H_{CM}\rangle$&Eigenvalues&Eigenvectors&Mass&$\Omega_{bc}\bar{B}_s$&$\Omega_{bb}D_s$\\
$bbcs\bar{s}$& $\frac12^-$ &$\left[\begin{array}{ccccc}24.6&28.5&-19.8&-12.0&-17.0\\28.5&-7.8&-12.0&-30.5&7.5\\-19.8&-12.0&25.1&17.5&-43.3\\-12.0&-30.5&17.5&21.1&31.0\\-17.0&7.5&-43.3&31.0&-75.4\end{array}\right]$&$\left[\begin{array}{c}76.6\\34.3\\10.4\\-26.8\\-106.7\end{array}\right]$&$\left[\begin{array}{c}(-0.55,-0.44,0.48,0.52,0.01)\\(0.14,0.12,0.76,-0.45,-0.44)\\(0.76,0.01,0.20,0.62,-0.03)\\(-0.27,0.87,0.20,0.27,0.23)\\(0.17,-0.16,0.33,-0.28,0.87)\end{array}\right]$&$\left[\begin{array}{c}12987\\12945\\12921\\12884\\12804\end{array}\right]$&$\left[\begin{array}{c}12440\\12397\\12374\\12336\\12256\end{array}\right]$&$\left[\begin{array}{c}12350\\12308\\12284\\12247\\12167\end{array}\right]$\\
& $\frac32^-$ &$\left[\begin{array}{cccc}14.8&-14.5&12.0&11.9\\-14.5&1.2&30.6&49.1\\12.0&30.6&18.5&-29.3\\11.9&49.1&-29.3&-29.4\end{array}\right]$&$\left[\begin{array}{c}43.3\\32.3\\16.8\\-87.4\end{array}\right]$&$\left[\begin{array}{c}(-0.01,0.75,0.62,0.25)\\(0.39,-0.36,0.66,-0.53)\\(0.90,0.04,-0.19,0.40)\\(0.21,0.56,-0.38,-0.71)\end{array}\right]$&$\left[\begin{array}{c}12954\\12943\\12928\\12823\end{array}\right]$&$\left[\begin{array}{c}12406\\12395\\12380\\12276\end{array}\right]$&$\left[\begin{array}{c}12317\\12306\\12290\\12186\end{array}\right]$\\
&$\frac52^-$ &47.1&47.1&1&12958&12410&12321\\\hline
\end{tabular}
\end{table}

For the $QQQn\bar{n}$ states, we do not present the isospin indices in the results because the isoscalar and isovector cases are degenerate. If the isoscalar $QQQn\bar{n}$ and $QQQs\bar{s}$ can mix, the resulting states would have different masses with the isovector $QQQn\bar{n}$. Here, we do not consider this possibility.

In fact, one may also use the $H_Q$-$H_q$ type threshold to estimate the pentaquark masses, where the hadron $H_q$ contains only light quarks and $H_Q$ contains heavy quarks. From previous investigations \cite{Wu:2017weo,Wu:2016gas,Zhou:2018pcv}, we have seen that the multiquark masses estimated with such thresholds can be treated as theoretical lower limits. This feature should be related with the attractions inside conventional hadrons that cannot be taken into account in the present model. Since no evidence for triply heavy baryons is reported, with this feature, we may conversely set upper limits for the masses of the conventional triply heavy baryons. The formula is
\begin{eqnarray}
M_{H_{QQQ}}\leq\Big[M_{ref}-\langle\widehat{H}_{CMI}\rangle_{ref}\Big]-\Big[M_{H_{q\bar{q}}}-\langle\widehat{H}_{CMI}\rangle_{H_{q\bar{q}}}\Big]
+\langle\widehat{H}_{CMI}\rangle_{H_{QQQ}},
\end{eqnarray}
where the value of $[M_{ref}-\langle\widehat{H}_{CMI}\rangle_{ref}]=[M-\langle\hat{H}_{CMI}\rangle]$ may be calculated with definition or read out from tables \ref{penta-massccc}-\ref{penta-massbbc}. The obtained upper limits can be different if one considers different systems, but they do not rely on the angular momentum if a system is given. Of course, the correctness of this conjecture needs to be confirmed by further measurements. From Eq. \eqref{b-m-CMI} and the masses in above tables, we may set $M_{\Omega_{ccc}}\leq (M_{\Xi_{cc}^*}+M_D-M_\omega)+\frac{16}{3}(C_{cc}-C_{cn}+C_{n\bar{n}}+3C_{c\bar{n}})\approx 5044$ MeV if we consider the $cccn\bar{n}$ system. If the $cccs\bar{s}$ system is used, we have $M_{\Omega_{ccc}}\leq (M_{\Omega_{cc}^*}+M_{D_s}-M_\phi)+\frac{16}{3}(C_{cc}-C_{cs}+C_{s\bar{s}}+3C_{c\bar{s}})\approx 4897$ MeV. The upper limits extracted with the $cccn\bar{s}$ and $cccs\bar{n}$ are 4975 MeV and 4989 MeV, respectively. One should adopt the smallest value 4897 MeV as the final limit. Similarly, from our parameters, we have
\begin{eqnarray}
M_{\Omega_{bbb}}&\leq&(M_{\Omega_{bb}^*}+M_{B_s}-M_\phi)+\frac{16}{3}(C_{bb}-C_{bs}+C_{s\bar{s}}+3C_{b\bar{s}})\approx14640\text{ MeV},\nonumber\\
M_{\Omega_{ccb}^*}&\leq&(M_{\Omega_{bc}^*}+M_{D_s}-M_\phi)+\frac{8}{3}(C_{bc}+C_{cc}-C_{cs}-C_{bs}+2C_{s\bar{s}}+6C_{c\bar{s}})\approx8082\text{ MeV},\nonumber\\
M_{\Omega_{bbc}^*}&\leq&(M_{\Omega_{bb}^*}+M_{D_s}-M_\phi)+\frac{16}{3}(C_{bc}-C_{bs}+C_{s\bar{s}}+3C_{c\bar{s}})\approx11314\text{ MeV},
\end{eqnarray}
and further $M_{\Omega_{ccb}}\leq8029$ MeV and $M_{\Omega_{bbc}}\leq11261$ MeV. These limits are much lower than the baryon masses obtained with Eq. \eqref{mass}. Most of the masses obtained in the literature (table \ref{QQQ-sum}) are consistent with such constraints. Of course, these upper limits rely on the input masses of the doubly heavy baryons, but are irrelevant with the pentaquark masses. In this sense, the masses of conventional triply heavy baryons are constrained by those of conventional doubly heavy baryons. On the other hand, if the experimentally observed triply heavy baryons have masses larger than such values, one may conclude that they probably are not ground $QQQ$ states.

\begin{figure}[!h]
\centering
\begin{tabular}{ccc}
\scalebox{0.6}{\includegraphics{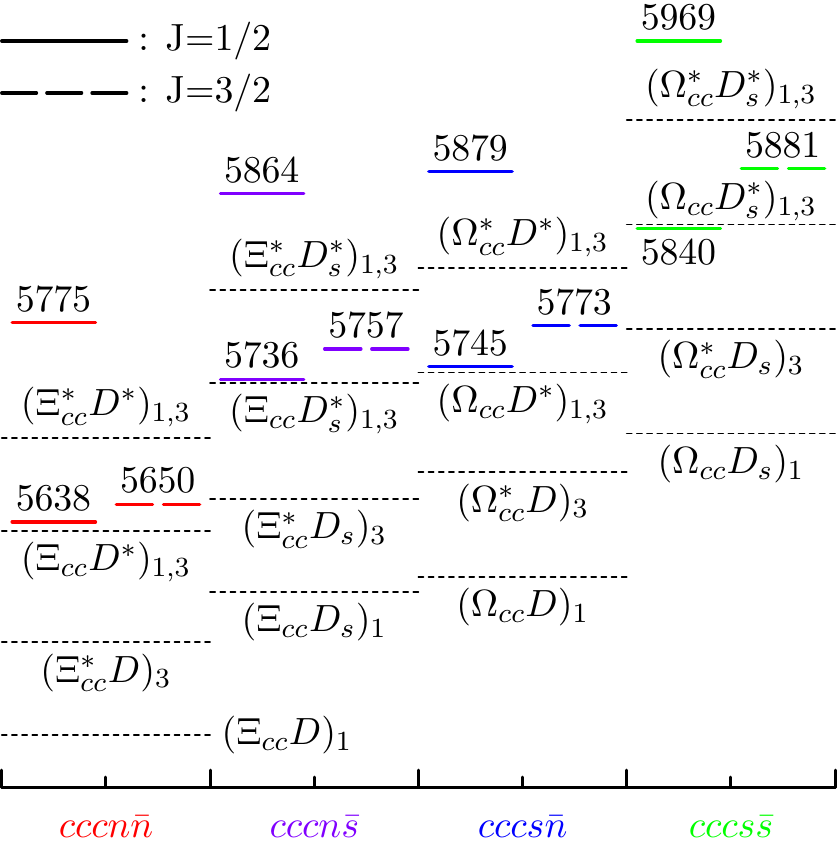}}\label{ccc}&\quad&\scalebox{0.6}{\includegraphics{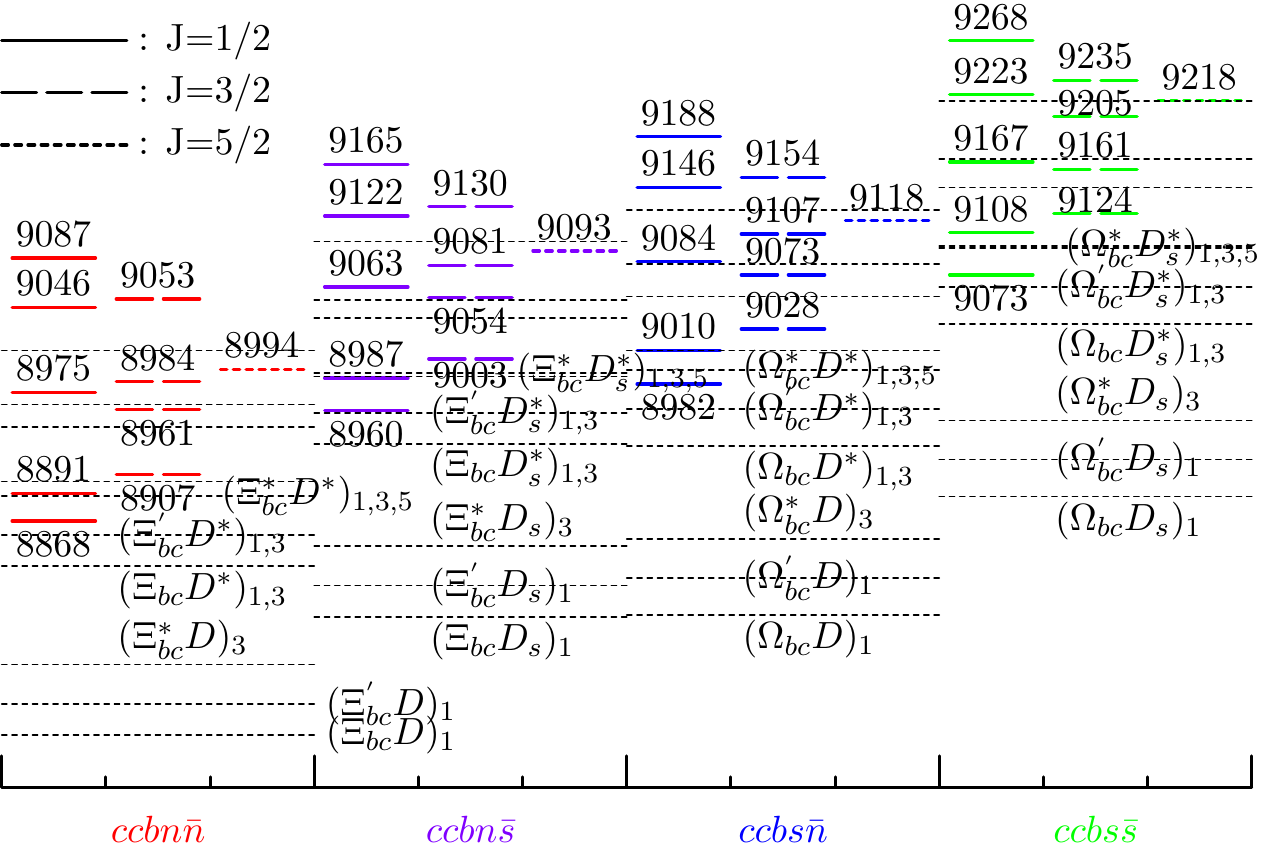}}\label{ccb}\\
(a) $cccq\bar{q}$ &&(c) $ccbq\bar{q}$\\
\scalebox{0.6}{\includegraphics{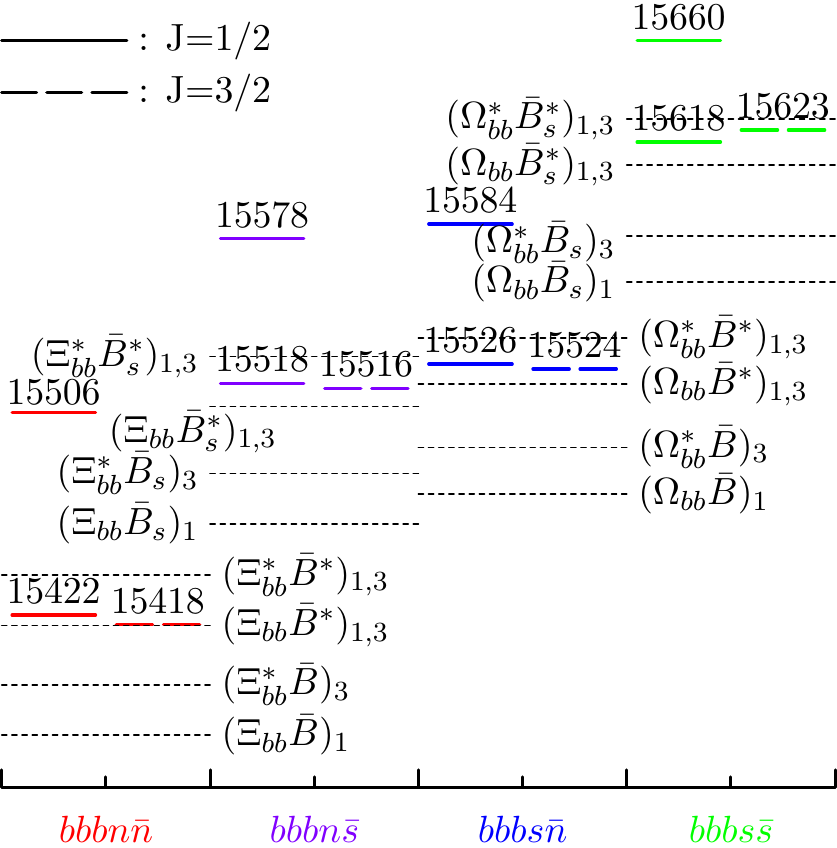}}\label{bbb}&\quad&\scalebox{0.6}{\includegraphics{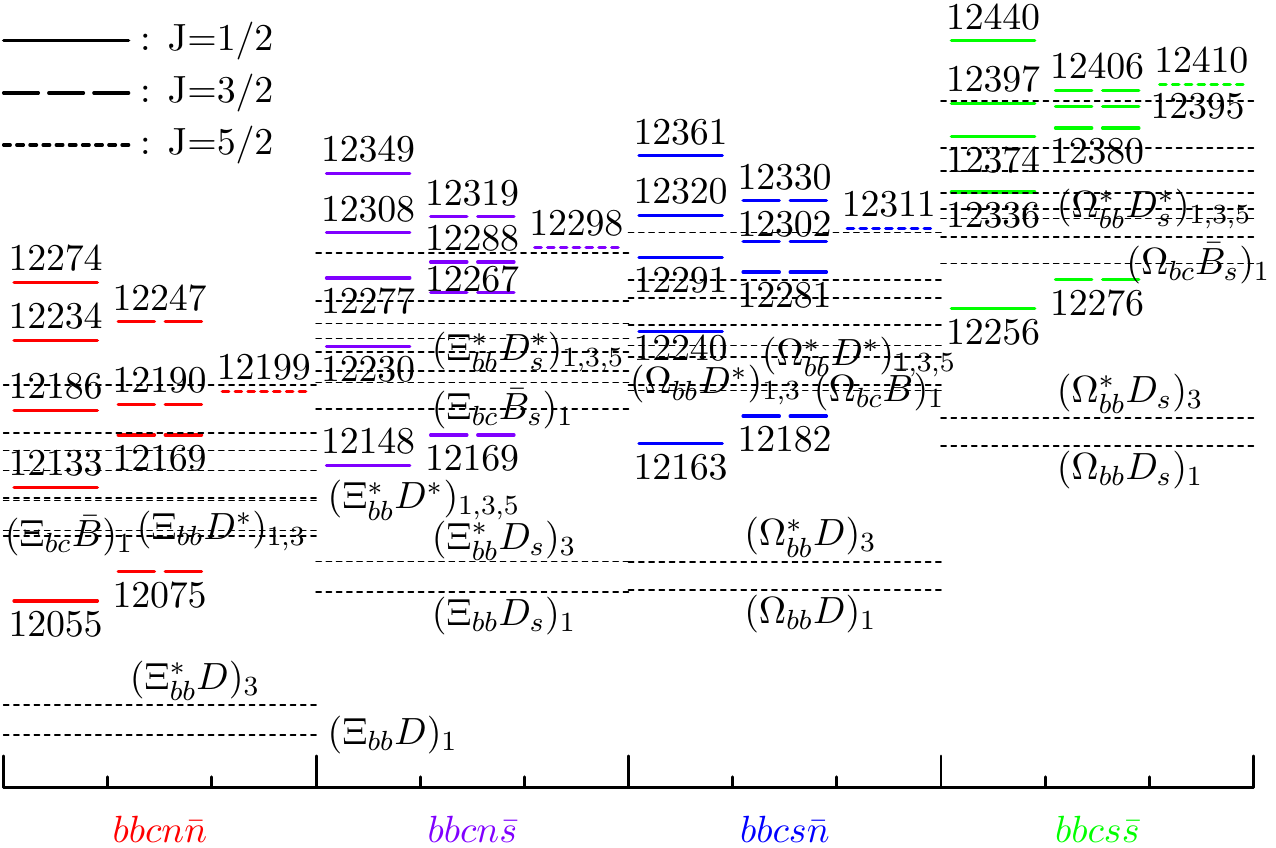}}\label{bbc}\\
(b) $bbbq\bar{q}$ &&(d) $bbcq\bar{q}$
\end{tabular}
\caption{Rough positions for the obtained triply heavy pentaquarks.}\label{fig}
\end{figure}

In Fig. \ref{fig}, we show the rough positions of the studied pentaquark states and the thresholds of relevant $(QQq)-(Q\bar{q})$, $(QQq)-(Q'\bar{q})$, or $(QQ'q)-(Q\bar{q})$ type decay patterns. Each diagram corresponds to a group of pentaquarks. We have adopted the masses listed in the seventh columns of tables \ref{penta-massccc}-\ref{penta-massbbc}. From a study of dibaryon states \cite{Park:2015nha}, the additional kinetic energy may be a possible effect in understanding why the predition of the $H$-dibaryon \cite{Jaffe:1976yi} is inconsistent with experimental results. If quark dynamics are considered, probably reasonable pentaquarks have higher masses than those obtained here. That is why we do not use the masses in the eighth columns of tables \ref{penta-massccb} and \ref{penta-massbbc} for the $QQQ^\prime q\bar{q}$ systems when plotting the diagrams. In Fig. \ref{fig}, the spectrum relies on the input masses of the doubly heavy baryons, but the relative positions comparing to the thresholds relating to the mass estimation is fixed in the present method. For example, the masses of the $J=5/2$ $ccbn\bar{n}$ states ($I=1$ and $I=0$) may be changed if one uses a different input mass of $\Xi_{bb}$ but the distance between the threshold of $\Xi_{bb}\bar{B}$ and them is fixed by the CMI matrices. Of course, the mass splittings between the pentaquark states are also fixed only by the CMI matrices.

From tables \ref{penta-massccc}-\ref{penta-massbbc} and Fig. \ref{fig}, one may understand some features for the spectra we study: i). There is a hierarchy around 3200 MeV between the four groups of pentaquarks, which is from the mass difference between the bottom and charm quarks. Within each group, the hierarchy caused by the mass difference $m_s-m_n$ is about 100 MeV; ii). For each system, the states with highest and lowest masses usually have the lowest spin $J=1/2$; iii). The $QQQs\bar{n}$ states generally have higher masses than the $QQQn\bar{s}$ states, which is a result of the difference in coupling strengths, $C_{Qn}\neq C_{Qs}$ and $C_{Q\bar{n}}\neq C_{Q\bar{s}}$.

If the considered compact pentaquarks exist, they will decay into lower hadron states. The easiest decay mode, $(QQq)+(Q\bar{q})$, $(QQq)+(Q'\bar{q})$, or $(QQ'q)+(Q\bar{q})$, would be due to quark rearrangements. Fig. \ref{fig} shows relevant thresholds for such a mode. When $2J$ ($J$ is the spin of a pentaquark) is equal to a subscript of the label for the baryon-meson channel, the decay into that channel through $S$-wave is allowed. The other decay mode is $(QQQ)+(q\bar{q})$ or $(QQQ')+(q\bar{q})$. Now, the color of the three heavy quarks is changed by emitting a gluon. Because of the possible constraints from the Pauli principle and/or angular momentum, the spin-flip of $QQQ$ may happen and this decay mode is suppressed for some cases where the spin-flip of $QQQ$ is forbidden in the heavy quark limit. We check this feature in the following analyses.

\subsection{Stability of various states}

We explore the possible compact structure $QQQq\bar{q}$ in this article, where $QQQ$ is always a color octet state. Since the mass estimation for the pentaquark states depends on effective quark masses or the masses of the unobserved doubly heavy baryons, what we obtain here are only the rough positions of such states. More accurate values need further dynamical calculations. Before those studies, it is beneficial to discuss preliminarily the stability of such pentaquark states and to find out possible interesting exotic states.

The basic idea is to consider the allowed two-body strong decays and the size of the phase space for decay. From the above results, all the pentaquark states seem to have rearrangement decays and are probably not stable. Since hadrons are composed of quarks, the masses and decay properties are finally determined by the quark-quark interactions. It is helpful to understand the hadron-level features from the inner interactions. If heavy quark spin symmetry plays a role in the decay processes \cite{Voloshin:2004mh,Liu:2013rxa,Voloshin:2018tav}, probably narrow pentaquarks are still possible. We will check this possibility. In the study of the $QQ\bar{Q}\bar{Q}$ systems \cite{Wu:2016vtq}, we checked the effective interaction between two heavy quarks in the mixing case by varying the effective coupling strengths, through which we may roughly guess whether the tetraquark states are stable or not. Here, we also perform a similar study.

\subsubsection{The $cccq\bar{q}$ and $bbbq\bar{q}$ states}

For these states, the spin of $ccc$ or $bbb$ ($QQQ$) is always 1/2 from the symmetry consideration, irrespective of the total spin of the system, 1/2 or 3/2. Since the spin of the color-singlet $\Omega_{ccc}$ or $\Omega_{bbb}$ is 3/2, the decay of such pentaquarks into a triply heavy baryon plus a light meson involves an emission of a chromomagnetic gluon and the spin-flip of $QQQ$. In the heavy quark limit, such a decay process is suppressed.
In principle, no symmetry principle suppresses the decay into a doubly heavy baryon plus a heavy-light meson and one expects that narrow $QQQq\bar{q}$ states would not exist if the pentaquark masses are high enough. To check this point, one may calculate overlapping factor between the initial state wave function and the final state wave function. From the recoupling formula in the spin space, for the $cccn\bar{n}$ case (other $QQQq\bar{q}$ cases are similar), one has
\begin{eqnarray}\label{finalwavefunction}
(\Xi_{cc}D)^{J=\frac12}&=&\frac12(ccc)^{\frac12}(n\bar{n})^{0}-\frac{1}{2\sqrt3}(ccc)^{\frac12}(n\bar{n})^1+\sqrt{\frac23}(ccc)^{\frac32}(n\bar{n})^{1},\nonumber\\
(\Xi_{cc}D^*)^{J=\frac12}&=&-\frac{1}{2\sqrt3}(ccc)^{\frac12}(n\bar{n})^{0}+\frac56(ccc)^{\frac12}(n\bar{n})^1+\frac{\sqrt2}{3}(ccc)^{\frac32}(n\bar{n})^{1},\nonumber\\
(\Xi_{cc}^*D^*)^{J=\frac12}&=&\sqrt{\frac23}(ccc)^{\frac12}(n\bar{n})^{0}+\frac{\sqrt2}{3}(ccc)^{\frac12}(n\bar{n})^1-\frac13(ccc)^{\frac32}(n\bar{n})^{1};\nonumber\\
(\Xi_{cc}^*D)^{J=\frac32}&=&-\frac{1}{\sqrt3}(ccc)^{\frac12}(n\bar{n})^{1}+\frac12(ccc)^{\frac32}(n\bar{n})^0+\frac{\sqrt{15}}{6}(ccc)^{\frac32}(n\bar{n})^{1},\nonumber\\
(\Xi_{cc}D^*)^{J=\frac32}&=&\frac13(ccc)^{\frac12}(n\bar{n})^{1}-\frac{1}{\sqrt3}(ccc)^{\frac32}(n\bar{n})^0+\frac{\sqrt5}{3}(ccc)^{\frac32}(n\bar{n})^{1},\nonumber\\
(\Xi_{cc}^*D^*)^{J=\frac32}&=&\frac{\sqrt5}{3}(ccc)^{\frac12}(n\bar{n})^{1}+\frac{\sqrt{15}}{6}(ccc)^{\frac32}(n\bar{n})^0+\frac16(ccc)^{\frac32}(n\bar{n})^{1};
\end{eqnarray}
where the superscripts indicate the spins. Obviously, there is no suppressed coupling from the comparison between these wave functions and those in Eqs. \eqref{Awavefunction} and \eqref{Bwavefunction}. One may roughly get relative coupling amplitudes between different channels with these wave functions, but could not derive the absolute partial widths without knowing the coupling strength. Therefore, the possible nature that the $QQQq\bar{q}$ states have generally broad widths cannot be excluded from this coupling analysis.

In a multiquark state, the interactions between each pair of quark components together affect the final mass which determines the phase space of decay. After the complicated coupling among different color-spin structures, the property of the interaction between two quark components becomes unclear. It is helpful to understand whether the effective interaction is attractive or repulsive by introducing a measure. To find it, we may study the effects induced by the artificial change of coupling strengths in the Hamiltonian. When a coupling strength is reduced, the resulting mass may be increased or decreased. If the effective interaction between the considered components is attractive (repulsive), the mass would be shifted upward (downward) and vice versa. For the purpose to see the effects, we may define a dimensionless variable
\begin{eqnarray}
K_{ij}=\frac{\Delta M}{\Delta C_{ij}},
\end{eqnarray}
where $\Delta C_{ij}$ is the variation of a coupling strength in Eq. \eqref{CMI} and $\Delta M=\Delta\langle\hat{H}_{CMI}\rangle$ is the corresponding variation of a pentaquark mass. Then the positive (negative) $K_{ij}$ corresponds to a repulsive (attractive) effective interaction between the $i$th and $j$th components. When $\Delta C_{ij}$ is small enough, $K_{ij}$ tends to be a constant $(\partial M)/(\partial C_{ij})$. With these constants, we may write an equivalent expression for the pentaquark mass as
\begin{eqnarray}\label{equivalentCMI}
M=M_0+\langle\hat{H}_{CMI}\rangle = M_0+\sum_{i<j}K_{ij}C_{ij},
\end{eqnarray}
where $M_0$ is a constant for a given pentaquark system and its value can be obtained based on the estimation approach, Eq. \eqref{mass} or \eqref{massref}.  One should note that this formula does not mean that $M$ is linearly related with $C_{ij}$.

\begin{table}[htbp]
\caption{Values of $K_{ij}$ when reducing the coupling strength $C_{ij}$ between the $i$th and $j$th quark components by 0.01\%. The orders of states are the same as those in tables \ref{penta-massccc} and \ref{penta-massbbb}.}\label{varQQQqqbar}
\begin{tabular}{ccccc|ccccc}\hline
State&$n\bar{n}$&$c\bar{n}$&$cn$&$cc$&State&$n\bar{n}$&$b\bar{n}$&$bn$&$bb$\\
$[cccn\bar{n}]^{J=\frac12}$&1.27&6.97&3.32&10.00&$[bbbn\bar{n}]^{J=\frac12}$&1.84&3.09&2.31&10.00\\
&0.06&-0.31&-9.99&10.00&&-0.51&3.58&-8.98&10.00\\
$[cccn\bar{n}]^{J=\frac32}$&-0.67&-3.33&3.33&10.00&$[bbbn\bar{n}]^{J=\frac32}$&-0.67&-3.33&3.33&10.00\\\hline

State&$n\bar{s}$&$c\bar{s}$&$cn$&$cc$&State&$n\bar{s}$&$b\bar{s}$&$bn$&$bb$\\
$[cccn\bar{s}]^{J=\frac12}$&1.00&8.09&3.09&10.00&$[bbbn\bar{s}]^{J=\frac12}$&1.63&4.93&3.07&10.00\\
&0.33&-1.42&-9.76&10.00&&-0.29&1.73&-9.74&10.00\\
$[cccn\bar{s}]^{J=\frac32}$&-0.67&-3.33&3.33&10.00&$[bbbn\bar{s}]^{J=\frac32}$&-0.67&-3.33&3.33&10.00\\\hline

State&$s\bar{n}$&$c\bar{n}$&$cs$&$cc$&State&$s\bar{n}$&$b\bar{n}$&$bs$&$bb$\\
$[cccs\bar{n}]^{J=\frac12}$&1.03&7.99&3.12&10.00&$[bbbs\bar{n}]^{J=\frac12}$&1.67&4.60&2.97&10.00\\
&0.31&-1.33&-9.79&10.00&&-0.34&2.06&-9.64&10.00\\
$[cccs\bar{n}]^{J=\frac32}$&-0.67&-3.33&3.33&10.00&$[bbbs\bar{n}]^{J=\frac32}$&-0.67&-3.33&3.33&10.00\\\hline

State&$s\bar{s}$&$c\bar{s}$&$cs$&$cc$&State&$s\bar{s}$&$b\bar{s}$&$bs$&$bb$\\
$[cccs\bar{s}]^{J=\frac12}$&0.69&9.04&2.51&10.00&$[bbbs\bar{s}]^{J=\frac12}$&0.99&8.14&3.07&10.00\\
&0.64&-2.37&-9.17&10.00&&0.35&-1.47&-9.74&10.00\\
$[cccs\bar{s}]^{J=\frac32}$&-0.67&-3.33&3.33&10.00&$[bbbs\bar{s}]^{J=\frac32}$&-0.67&-3.33&3.33&10.00\\\hline
\end{tabular}
\end{table}

Now we apply this definition of $K_{ij}$ to the $cccn\bar{n}$ states. By reducing one relevant coupling strength to 99.99\% of its original value, we get a $K_{ij}$. Each time we change only one $C_{ij}$ while others remain unchanged. The results are collected in table \ref{varQQQqqbar} where the order of states is the same as that in table \ref{penta-massccc}. A positive (negative) number indicates that the corresponding color-spin interaction is effectively repulsive (attractive). In the $bc\bar{b}\bar{c}$ or $bb\bar{b}\bar{b}$ case, for example, one may understand that a relatively stable tetraquark is favored if the effective $bc$ or $bb$ interaction is attractive. In the present case, not only one quark-quark interaction exists. Usually, it is hard to guess whether a pentaquark state is stable or not just from the signs of $K_{ij}$. The sum of their contributions gives the final mass shift. From table \ref{varQQQqqbar}, it is obvious that the effective color-spin interactions between each pair of components in the high mass $J=1/2$ $cccn\bar{n}$ states ($I=1$ and $I=0$) are repulsive and the states should not be stable. By checking the CMI matrix in Eq. \eqref{CMI1} where the interaction between quark components can be attractive, one understands that the mixing effect may change the nature of each interaction. In addition to judge the stability of a pentaquark with Fig. \ref{fig}, one may equivalently use Eq. \eqref{equivalentCMI} to judge whether the decay into a $(ccn)-(c\bar{n})$ channel happens or not. Since $K_{cc}C_{cc}+K_{cn}C_{cn}>8/3C_{cc}+16/3C_{cn}$ (CMI of $\Xi_{cc}^*$) and $K_{c\bar{n}}C_{c\bar{n}}>16/3C_{c\bar{n}}$ (CMI of $D^*$), the decay into $\Xi_{cc}^*D^*$ (and thus $\Xi_{cc}D^*$ and $\Xi_{cc}D$) through $S$-wave is allowed. If this state exists, its width should not be narrow. For a $J=3/2$ state ($I=1$ or $I=0$), the $n\bar{n}$ and $c\bar{n}$ interactions are both attractive, which is obvious from Eq. \eqref{CMI2} or \eqref{equivalentCMI}, and the width of this state should be relatively narrower. From masses or effective interactions, $S$-wave $\Xi_{cc}D^*$ and $\Xi_{cc}^*D$ decay channels are both opened. For the remaining $J=1/2$ states, the attraction appears mainly between a charm quark and a light quark. Now, the allowed $S$-wave decay channels are $\Xi_{cc}D$ and possibly $\Xi_{cc}D^*$. The same $K_{cc}$ for different states reflects the fact that the spin of $ccc$ is always 3/2. From these discussions, it seems that all these $cccn\bar{n}$ pentaquarks are not narrow states even if they do exist.

For the other $cccq\bar{q}$ states, only one isospin is allowed, either 1/2 or 0. The higher $J=1/2$ $cccn\bar{s}$ state has three $S$-wave channels $\Xi_{cc}^*D_s^*$, $\Xi_{cc}D_s^*$, and $\Xi_{cc}D_s$ and the $J=3/2$ state has two $S$-wave channels $\Xi_{cc}D_s^*$ and $\Xi_{cc}^*D_s$. The allowed $S$-wave decay channels for the remaining $J=1/2$ $cccn\bar{s}$ state are $\Xi_{cc}D_s$ and possibly $\Xi_{cc}D_s^*$. Effective contributions from each pair of quark components can also be obtained from the values shown in table \ref{varQQQqqbar}. In principle, these $cccn\bar{s}$ states should not be narrow. The $cccs\bar{n}$ states have similar decay properties, but $D_s$ ($D_s^*$) is replaced by $D$ ($D^*$) and $\Xi_{cc}$ ($\Xi_{cc}^*$) is replaced by $\Omega_{cc}$ ($\Omega_{cc}^*$). If one replaces only $\Xi_{cc}$ ($\Xi_{cc}^*$) by $\Omega_{cc}$ ($\Omega_{cc}^*$) in the decay products of the $cccn\bar{s}$ states, the decay patterns of the $cccs\bar{s}$ states are obtained. From the above discussions, the widths of all the $cccq\bar{q}$ should not be narrow.

Similarly, one gets rough decay properties of the $bbbq\bar{q}$ states. It is easy to understand that their widths should not be narrow either, since all of them have $S$-wave rearrangement decay patterns and no suppressed channels are found.\\

If the interactions between light quark components have large contributions to the pentaquark mass, the state may be low enough and probably stable, because the coupling constant in the CMI model is proportional to $1/(m_im_j)$. Two examples are the $udsc\bar{c}$ state proposed in Refs. \cite{Wu:2017weo,Irie:2017qai} and the $QQqq\bar{q}$ states studied in Ref. \cite{Zhou:2018pcv}. In the present $QQQq\bar{q}$ case, it is easy to read out contributions from each pair of interaction from table \ref{varQQQqqbar}. Obviously, the $q\bar{q}$ effective interactions are not attractive enough (even repulsive). The effective interactions for heavy-heavy and heavy-light components do not provide enough attraction, either. Thus, the masses of $QQQq\bar{q}$ states are not low enough, which results in the conclusion that such pentaquark states are probably not narrow even if they exist. Of course, further conclusion needs accurate hadron masses and the determination of decay amplitudes.

\subsubsection{The $ccbq\bar{q}$ and $bbcq\bar{q}$ states}

Without the constraint from the Pauli principle, the spin of $ccb$ or $bbc$ ($QQQ'$) can be both 3/2 and 1/2. The resulting spectrum is more complicated. When the angular momentum of the pentaquark is 5/2, the spin of $QQQ'$ must be 3/2. The decay of the pentaquark into $\Omega_{QQQ'}$ plus a light meson is forbidden or suppressed, but the decay into $\Omega_{QQQ'}^*$ plus a light vector meson not. The latter decay is forbidden only by kinematics. When the angular momentum of the pentaquark is 1/2 or 3/2, the spin of the inside $QQQ'$ can be both 1/2 and 3/2. From the eigenvectors in tables \ref{penta-massccb} and \ref{penta-massbbc}, the spin of $QQQ'$ in some states is dominantly 1/2. The decays of the 1/2 or 3/2 pentaquarks into $\Omega_{QQQ'}$ and a $q\bar{q}$ meson are not suppressed, either. For the $(QQq)+(Q'\bar{q})$ decay mode, it is easy to understand that each overlapping factor does not vanish from Eqs. \eqref{Cwavefunction}, \eqref{Dwavefunction}, and \eqref{finalwavefunction}. (The nonvanishing overlapping factor for the decay of a $J=5/2$ pentaquark is obvious.) For the $(QQ'q)+(Q\bar{q})$ decay mode, the nonvanishing overlapping factor can be understood similarly since the spin of $QQ'$ in a physical $(QQ'q)$ state can be both 1 and 0. Even if the spin of $QQ'$ in $(QQ'q)$ is only zero, the factor still does not vanish. Therefore, no symmetry requirement can suppress the decay of the $QQQ'q\bar{q}$ states. Whether such a state can decay or not depends only on kinematics.

\begin{table}[htbp]
\caption{Values of $K_{ij}$ when reducing the coupling strength $C_{ij}$ between the $i$th and $j$th quark components by 0.001\%. The orders of states are the same as those in tables \ref{penta-massccb} and \ref{penta-massbbc}.}\label{varQQQpqqbar}
\begin{tabular}{cccccccc|cccccccc}\hline
State&$n\bar{n}$&$c\bar{n}$&$b\bar{n}$&$cn$&$bn$&$cc$&$bc$&State&$n\bar{n}$&$c\bar{n}$&$b\bar{n}$&$cn$&$bn$&$bb$&$bc$\\
$[ccbn\bar{n}]^{J=\frac12}$&1.45&4.44&1.52&1.85&1.40&3.38&6.57&$[bbcn\bar{n}]^{J=\frac12}$&1.65&2.13&2.49&0.64&2.28&3.28&6.65\\
&1.52&4.35&1.93&1.65&-3.16&3.35&-4.45&&1.71&2.38&1.80&-0.05&0.33&3.28&-4.59\\
&0.02&1.43&-1.74&-6.69&-2.47&3.30&5.57&&-0.29&3.06&-2.77&-0.82&-7.53&3.01&4.70\\
&-0.57&-7.83&3.23&-2.76&-0.25&3.14&-4.78&&-0.60&1.30&-2.41&-4.82&-1.12&3.72&-3.54\\
&-0.42&-6.39&-8.26&-8.05&-2.19&2.83&-0.91&&-0.47&-12.21&-3.12&-1.61&-7.96&2.72&-1.22\\
$[ccbn\bar{n}]^{J=\frac32}$&1.47&3.61&0.14&2.51&0.45&2.88&-1.21&$[bbcn\bar{n}]^{J=\frac32}$&1.78&3.43&0.75&-1.29&0.44&2.67&-0.75\\
&-0.59&0.26&-3.11&2.90&-1.26&3.05&5.43&&-0.63&3.81&-3.13&1.45&-2.41&2.79&3.33\\
&-0.27&3.09&1.72&-6.00&1.62&3.00&-3.55&&-0.66&-0.35&3.89&2.27&-2.05&3.82&-1.14\\
&-0.61&-6.96&-4.09&0.59&0.52&3.07&-0.66&&-0.48&-12.23&-1.51&-1.09&4.02&2.73&-1.44\\
$[ccbn\bar{n}]^{J=\frac52}$&-0.67&1.33&4.67&4.67&1.33&2.67&-0.67&$[bbcn\bar{n}]^{J=\frac52}$&-0.67&4.67&1.33&1.33&4.67&2.67&-0.67\\\hline

State&$n\bar{s}$&$c\bar{s}$&$b\bar{s}$&$cn$&$bn$&$cc$&$bc$&State&$n\bar{s}$&$c\bar{s}$&$b\bar{s}$&$cn$&$bn$&$bb$&$bc$\\
$[ccbn\bar{s}]^{J=\frac12}$&1.14&5.26&2.22&1.98&1.25&3.36&6.63&$[bbcn\bar{s}]^{J=\frac12}$&1.36&2.53&3.94&0.84&2.45&3.31&6.66\\
&1.32&5.25&1.45&1.34&-3.81&3.45&-4.00&&1.41&3.15&2.09&-0.18&-1.15&3.07&-3.30\\
&0.39&0.23&-1.69&-6.76&-1.87&3.17&5.26&&0.18&2.34&-4.37&-0.88&-6.15&3.17&3.24\\
&-0.55&-8.08&2.89&-2.19&-0.41&3.17&-4.86&&-0.55&1.14&-2.54&-5.04&-0.96&3.72&-3.30\\
&-0.30&-6.66&-8.20&-8.37&-1.83&2.85&-1.03&&-0.40&-12.50&-3.11&-1.42&-8.19&2.72&-1.31\\
$[ccbn\bar{s}]^{J=\frac32}$&1.10&4.75&0.02&2.90&0.84&3.03&-1.57&$[bbcn\bar{s}]^{J=\frac32}$&1.62&3.91&0.82&-1.50&1.47&2.68&-0.80\\
&-0.57&-0.08&-2.69&3.43&-1.47&2.96&5.29&&-0.56&3.82&-2.57&1.43&-3.59&2.77&2.96\\
&0.03&2.58&2.28&-7.59&1.28&2.92&-2.86&&-0.65&-0.42&3.47&2.20&-2.07&3.82&-0.71\\
&-0.56&-7.24&-4.94&1.27&0.68&3.09&-0.85&&-0.40&-12.65&-1.72&-0.80&4.18&2.73&-1.45\\
$[ccbn\bar{s}]^{J=\frac52}$&-0.67&1.33&4.67&4.67&1.33&2.67&-0.67&$[bbcn\bar{s}]^{J=\frac52}$&-0.67&4.67&1.33&1.33&4.67&2.67&-0.67\\\hline

State&$s\bar{n}$&$c\bar{n}$&$b\bar{n}$&$cs$&$bs$&$cc$&$bc$&State&$s\bar{n}$&$c\bar{n}$&$b\bar{n}$&$cs$&$bs$&$bb$&$bc$\\
$[ccbs\bar{n}]^{J=\frac12}$&1.16&5.16&2.26&2.07&1.17&3.34&6.65&$[bbcs\bar{n}]^{J=\frac12}$&1.38&2.43&3.92&0.93&2.37&3.33&6.65\\
&1.33&5.22&1.33&1.33&-3.79&3.48&-3.95&&1.40&3.17&1.84&-0.13&-1.32&3.05&-3.08\\
&0.39&0.10&-1.51&-6.46&-1.96&3.20&5.04&&0.16&2.61&-4.86&-0.47&-6.05&3.13&2.63\\
&-0.59&-7.74&2.24&-2.19&-0.62&3.14&-4.75&&-0.52&0.86&-1.72&-5.51&-0.90&3.76&-2.87\\
&-0.29&-6.73&-7.66&-8.75&-1.47&2.85&-0.99&&-0.42&-12.41&-3.18&-1.48&-8.10&2.73&-1.33\\
$[ccbs\bar{n}]^{J=\frac32}$&1.10&4.75&-0.20&2.97&0.90&3.03&-1.53&$[bbcs\bar{n}]^{J=\frac32}$&1.64&3.84&0.72&-1.46&1.43&2.68&-0.76\\
&-0.59&-0.01&-2.93&3.70&-1.49&2.91&5.16&&-0.57&3.72&-2.85&1.46&-3.33&2.80&3.24\\
&0.09&2.41&1.89&-7.36&1.32&2.99&-3.05&&-0.65&-0.27&3.71&2.21&-2.27&3.81&-1.09\\
&-0.60&-7.15&-4.09&0.69&0.61&3.07&-0.58&&-0.42&-12.62&-1.58&-0.88&4.17&2.72&-1.39\\
$[ccbs\bar{n}]^{J=\frac52}$&-0.67&1.33&4.67&4.67&1.33&2.67&-0.67&$[bbcs\bar{n}]^{J=\frac52}$&-0.67&4.67&1.33&1.33&4.67&2.67&-0.67\\\hline

State&$s\bar{s}$&$c\bar{s}$&$b\bar{s}$&$cs$&$bs$&$cc$&$bc$&State&$s\bar{s}$&$c\bar{s}$&$b\bar{s}$&$cs$&$bs$&$bb$&$bc$\\
$[ccbs\bar{s}]^{J=\frac12}$&0.70&6.03&2.97&1.92&0.55&3.33&6.60&$[bbcs\bar{s}]^{J=\frac12}$&0.76&2.99&5.82&1.08&1.55&3.33&6.61\\
&1.14&6.03&0.40&0.52&-3.99&3.57&-3.19&&0.91&4.15&-0.00&-0.61&-4.61&2.71&0.28\\
&0.84&-1.12&-0.90&-5.89&-1.41&3.04&4.62&&0.98&1.81&-5.16&0.12&-1.91&3.43&-1.34\\
&-0.57&-7.86&1.73&-1.81&-0.76&3.15&-4.73&&-0.37&0.44&-1.51&-6.05&-0.66&3.78&-2.06\\
&-0.10&-7.08&-7.52&-8.74&-1.06&2.91&-1.30&&-0.29&-12.72&-3.14&-1.21&-8.37&2.74&-1.48\\
$[ccbs\bar{s}]^{J=\frac32}$&0.53&5.96&-0.57&2.74&1.51&3.25&-1.88&$[bbcs\bar{s}]^{J=\frac32}$&0.34&4.64&-1.35&-0.43&4.58&2.67&0.57\\
&-0.58&-0.18&-2.34&4.12&-1.61&2.83&4.66&&0.50&3.09&0.24&0.40&-5.84&2.87&1.41\\
&0.51&1.88&3.06&-8.73&0.49&2.82&-1.91&&-0.57&0.05&2.95&1.82&-3.10&3.74&-0.58\\
&-0.46&-7.66&-5.48&1.88&0.95&3.11&-0.87&&-0.28&-13.11&-1.84&-0.46&4.36&2.72&-1.40\\
$[ccbs\bar{s}]^{J=\frac52}$&-0.67&1.33&4.67&4.67&1.33&2.67&-0.67&$[bbcs\bar{s}]^{J=\frac52}$&-0.67&4.67&1.33&1.33&4.67&2.67&-0.67\\\hline
\end{tabular}
\end{table}

To understand the effective quark-quark or quark-antiquark color-spin interactions, we list $K_{ij}$ of each pair of interactions in table \ref{varQQQpqqbar} where $\Delta C_{ij}=0.00001 C_{ij}$. Contributions to the pentaquark mass from each pair of components are easy to obtain. For example, the effective $cc$ or $bb$ interaction is always repulsive. From the table, the interactions in all the highest states ($J=1/2$) are repulsive and such states are certainly not stable. For the lowest states ($J=1/2$) in each system, the chromomagnetic interactions except for $cc$ or $bb$ are all attractive. To understand whether the interactions result in low enough and thus possible narrow pentaquarks, what we need is to check whether the lowest states with various spins can decay.

First, we focus on the lowest $ccbn\bar{n}$ states. The thresholds of the decay mode containing a $(bcn)$ baryon are generally lower than those of the mode containing a $(ccn)$ baryon. For the case $J=5/2$, the states ($I=1$ and $0$) probably do not have the $S$-wave decay pattern $\Xi_{cc}^*B^*$, but they may decay into $\Xi_{bc}^*D^*$ and $\Omega_{ccb}^*\rho$ or $\Omega_{ccb}^*\omega$ through $S$-wave. Their widths should not be narrow. For the lowest $J=3/2$ states ($I=1$ and $0$), the channels $\Xi_{bc}D^*$, $\Xi_{bc}^*D$, and $\Xi_{bc}^\prime D^*$ are all opened. They can also decay into $\Omega_{ccb}^*\pi$, $\Omega_{ccb}\rho$, or $\Omega_{ccb}\omega$ through $S$-wave. Their widths should not be narrow, either.  For the lowest $J=1/2$ states, the decay patterns $\Xi_{bc}D^*$, $\Xi_{bc}D$, $\Xi_{bc}^\prime D$, $\Omega_{ccb}\pi$, $\Omega_{ccb}\omega$, etc. are all allowed, which should result in broad pentaquark states if they exist. Therefore, narrow $ccbn\bar{n}$ pentaquarks are not expected. However, this conclusion depends on the pentaquark masses. If the masses estimated with the $\Xi_{bc}D$ are more reasonable, one may conclude that relatively narrow $J=5/2$ or $J=3/2$ pentaquark states are still possible.

Next, one may analyse similarly the cases $ccbn\bar{s}$, $ccbs\bar{n}$, and $ccbs\bar{s}$. The basic conclusion is that relatively narrow $J=5/2$ pentaquarks are possible only when their masses are low enough.

Finally, the $bbcq\bar{q}$ systems have similar spectra to the $ccbq\bar{q}$ systems, but the thresholds of the decay mode containing a $bbq$ baryon are lower than those of the mode containing a $bcq$ baryon. If pentaquark masses shown in Fig. \ref{fig} are reasonable, no narrow or stable $bbcq\bar{q}$ states should exist.\\

From the above results, one basically finds that all the $QQQ^\prime q\bar{q}$ pentaquark states are probably not stable, although relatively narrow states cannot be excluded completely. The color-spin interaction itself does not lead to enough attraction. Further theoretical and experimental investigations on doubly heavy and triply heavy states may provide more information.

\subsection{Discussions}

In obtaining the rough masses of the studied pentaquark states, we have several uncertainties. First, the values of coupling constants $C_{ij}$ in the model Hamiltonian are determined from the conventional hadrons. Whether they can be used to multiquark states is still an open question, as noticed in Refs. \cite{Stancu:2009ka,Maiani:2014aja}. Secondly, the mass shifts from CMI energies for a system and thus the pentaquark masses are not well determined. If one uses the effective quark masses determined from conventional hadrons, one obtains hadron masses with overestimated values. To reduce the uncertainty, we phenomenologically adopt a reference hadron-hadron system for our purpose. Whether the selected system is appropriate or not is also a question to be answered. At present, we tend to use a system with high threshold in order to include possible additional effects like the additional kinetic energy \cite{Park:2015nha}. The uncertainty for the obtained pentaquark masses in this method will be checked by the future experimental measurements. Thirdly, the adopted model does not involve dynamics and much information is hidden in constant parameters. The parameters should be different from system to system. However, as a preliminary work for multiquark system where the few-body problem is difficult to deal with, the present investigation on basic features of multiquark spectra is still helpful for further theoretical and experimental studies on multiquark properties. We want to emphasize that the estimated masses do not indicate that all of these states should be bound. Whether such states exist or not needs the experimental measurements to confirm. On the other side, if one compact pentaquark state could be observed, the masses of its partner states may be predicted with our results.

If the pentaquark masses shown in Fig. \ref{fig} are all reasonable, a question arises to distinguish a compact multiquark state from a molecule state. In understanding the nature of the $P_c(4380)$ state, both the $(cqq)(\bar{c}q)$ molecule configuration and the $(c\bar{c})_{8_c}(qqq)_{8_c}$ pentaquark configuration are not contradicted with experimental measurements. The masses of the proposed $udsc\bar{c}$ in these two configurations are also consistent \cite{Wu:2010jy,Wu:2017weo}. For the present $QQQq\bar{q}$ case, the situation is different. Our study does not favor low mass $cccq\bar{q}$ or $bbbq\bar{q}$ pentaquarks while the investigations at the hadron level \cite{Chen:2017jjn,Guo:2013xga} indicate that the molecule states such as $\Xi_{cc}D$ and $\Xi_{cc}D^*$ are possible. A study with the QCD sume rule method also favors the existence of such molecules \cite{Azizi:2018dva}. If a low-lying pentaquark-like state around such thresholds were observed, the molecule interpretation would be preferred. However, for the possible states around the thresholds of $\Xi_{cc}\bar{B}$ and $\Xi_{cc}\bar{B}^*$, one needs to use other information like branching ratios to understand whether it is a compact pentaquark or a molecule, which can be studied in future works.

As mentioned in the beginning of this section, we have used the assumptions that $C_{cc}=C_{c\bar{c}}$, $C_{s\bar{s}}=C_{ss}$, $C_{bb}=C_{b\bar{b}}$, and $C_{bc}=C_{b\bar{c}}$ because of lack of experimental data. This assumption certainly leads to uncertainties for the estimated masses. To see the differences caused by assumption selection, one may try other methods in determining the coupling parameters and compare the resulting masses. First, we check the extraction of $C_{cc}$, $C_{bb}$, and $C_{bc}$ from baryon masses. Based on the formulas given in Eq. \eqref{b-m-CMI}, one cannot determine $C_{cc}$ just from the mass difference between $\Xi_{cc}$ and $\Xi_{cc}^*$, but may get a number from $(M_{\Xi_{cc}}+2M_{\Xi_{cc}^*}-M_N-3M_{\eta_c}-8C_{nn}-48C_{c\bar{c}})/8$, $3/8(M_{\Sigma}+M_{\Omega_{cc}}-M_{\Sigma_c}-M_{\Omega_c})+C_{ss}+4(C_{ns}-C_{cn})$, or $3/8(M_\Sigma+M_{\Omega_{cc}}-M_{\Xi_c^\prime}-M_{\Xi_c^*})+6C_{ns}+3C_{cs}-C_{nn}-C_{cn}$ once the mass of another doubly charmed baryon is known. However, the obtained $C_{cc}$ is either negative or unreasonably large if $M_{\Xi_{cc}^*}>M_{\Xi_{cc}}=3621.4$ MeV or $M_{\Omega_{cc}}<3850$ MeV (see Ref. \cite{Wei:2015gsa} for the rough value of $M_{\Omega_{cc}}$ in theoretical investigations). One may also use $(M_\Delta+2M_{\Omega_{ccc}}-2M_{\Xi_{cc}^*}-M_{\Xi_{cc}}-8C_{nn})/8$ to extract $C_{cc}$, but the reasonable (positive, not large) value is sensitive to the assignment for the spin of $\Xi_{cc}$ and the masses of the involved baryons. To extract $C_{bb}$, the masses of two doubly bottom baryons are at least needed, but similar situation to $C_{cc}$ is found once quark model predictions are used. For the value of $C_{bc}$, one may extract it with $(M_{\Omega_{ccb}^*}-M_{\Omega_{ccb}})/16=(M_{\Omega_{bbc}^*}-M_{\Omega_{bbc}})/16$ or from $M_{\Xi_{bc}^*}-M_{\Xi_{bc}^\prime}$. If the mass difference $M_{\Omega_{ccb}^*}-M_{\Omega_{ccb}}$ is around tens of MeV, the obtained $C_{bc}$ is not far from the adopted 3.3 MeV. However, $C_{bc}$ is sensitive to the mass difference $M_{\Xi_{bc}^*}-M_{\Xi_{bc}^\prime}$. One may also extract $m_c$, $m_b$, $C_{cc}$, $C_{bb}$, and $C_{bc}$ simultaneously by using a set of masses for the triply heavy baryons shown in Table \ref{QQQ-sum}, but results sensitive to baryon masses or unreasonable results are obtained. The above considerations indicate that the determination of $C_{cc}$, $C_{bc}$, and $C_{bb}$ with baryon masses still has some problems. Secondly, we check the method $C_{ij}=C_H/(m_im_j)$ usually used in the literature, e.g. Refs. \cite{Lee:2007tn,Lee:2009rt,Kim:2014ywa}. This method allows us to estimate $C_{s\bar{s}}$ from $C_{n\bar{n}}$. According to Ref. \cite{Lee:2007tn}, we have $C_{s\bar{s}}=9/25C_{n\bar{n}}=10.7$ MeV. This value is comparable to that from Ref. \cite{Maiani:2004vq}, $3/8\kappa_{s\bar{s}}=11.3$ MeV, but larger than that we have used. However, for $C_{cc}$, $C_{bb}$, and $C_{bc}$, the parameter $C_H$ is different from those for quark-antiquark pairs and light diquarks \cite{Lee:2007tn}. The determination of $C_H$ in this case needs more baryon masses and the above problems are probably also be there. Thirdly, we check the estimation method $C_{ij}=C_{i\bar{j}}C_{nn}/C_{n\bar{n}}\approx 2/3C_{i\bar{j}}$ where $C_{i\bar{j}}$ means $C_{ij}$ for a quark-antiquark pair. By inspecting values we haved adopted, those we can determine from experimental data are consistent with this method. Then one gets $C_{s\bar{s}}=10.5$ MeV, $C_{cc}=3.3$ MeV, $C_{bb}=1.8$ MeV, and $C_{bc}=2.0$ MeV. The obtained $C_{s\bar{s}}$ in this method is also consistent with that in the second method. We will compare estimated $QQQq\bar{q}$ masses using these parameters with those using our assumptions.

By changing $C_{cc}=C_{c\bar{c}}=5.3$ MeV, $C_{bb}=C_{b\bar{b}}=2.9$ MeV, $C_{bc}=C_{b\bar{c}}=3.3$ MeV, and $C_{s\bar{s}}=C_{ss}=6.5$ MeV to 3.3 MeV, 1.8 MeV, 2.0 MeV, and 10.5 MeV, respectively, and re-calculating pentaquark masses, one finds that this does not affect the results significantly: (1) the reduction of $C_{cc}$ or $C_{bb}$ results in $\sim$10 MeV lower masses for relevant systems and the effects caused by the change of $C_{bc}$ or $C_{s\bar{s}}$ are just around several MeV's; and (2) the change of all the coupling parameters simultaneously induces mass shifts around 0$\sim$20 MeV. In fact, the effects due to the changes of $C_{ij}$ can be roughly seen from the obtained $K_{ij}$'s although the value of $K_{ij}$ is also affected by the coupling parameters. From tables \ref{varQQQqqbar} and \ref{varQQQpqqbar}, the largest numbers are $K_{cc}=K_{bb}=10$. Since the variations of $C_{cc}$, $C_{bb}$, and $C_{bc}$ cannot be large, the mass uncertainties because of them are at most tens of MeV. For $C_{s\bar{s}}$, the effects from its uncertainty are not large, either, because the largest $K_{s\bar{s}}$ is only around 1. Therefore, from the above discussions, the approximations we have adopted do not cause significant effects on the results.

Now we move on to the question how to distinguish an orbitally excited $QQQ$ structure from an isoscalar pentaquark structure if a high mass $QQQ$-like state was observed. Since the conventional $P$-wave states and the pentaquark states may have similar decay patterns, the theoretical mass gaps between these two structures are useful. We can check the gaps between the upper limits for the masses of ground $QQQ$ baryons and the lowest $(QQn)(Q\bar{n})$ thresholds ($\Xi_{cc}D\sim$5490 MeV, $\Xi_{bb}\bar{B}\sim$15370 MeV, $\Xi_{bc}D\sim$8690 MeV, and $\Xi_{bb}D\sim$11960 MeV) above which pentaquarks probably exist. They are around 590 MeV for the $ccc$ case, 730 MeV for the $bbb$ case, 600 MeV for the $ccb$ case, and 650 MeV for the $bbc$ case. According to the calculations in the literature, the $P$-wave excitation energy for the $ccc$ case is around 120$\sim$360 MeV \cite{Wang:2011ae,Roberts:2007ni,Aliev:2014lxa,Padmanath:2013zfa}, that for the $bbb$ case is around 120$\sim$600 MeV \cite{Wang:2011ae,Roberts:2007ni,Aliev:2014lxa,Meinel:2012qz}, that for the $ccb$ case is around 130$\sim$280 MeV \cite{Wang:2011ae,Roberts:2007ni,Aliev:2014lxa}, and that for the $bbc$ case is around 120$\sim$250 MeV \cite{Wang:2011ae,Gershtein:1998un,Roberts:2007ni,Aliev:2014lxa,Shah:2018div}. Therefore, the $P$-wave $QQQ$ states are expected to be below the $(QQn)(Q\bar{n})$ thresholds and thus below the pentaquarks. Alternatively, in theoretical studies, the maximum masses of $P$-wave states are \cite{Roberts:2007ni}: $(ccc)\sim 5160$ MeV, $(bbb)\sim 14976$ MeV, $(ccb)\sim8432$ MeV, and $(bbc)\sim 11762$ MeV. It seems that one does not need to worry about the exotic nature of the observed $QQQ$ baryon once it is below relevant $(QQn)(Q\bar{n})$ thresholds. However, this argument is only applicable to the lowest $P$-wave $QQQ$ states. If the $P$-wave $QQQ$ states (e.g. radially excited states studied in Ref. \cite{Shah:2018div}) are heavier, the coupled channel effects due to the $(QQn)(Q\bar{n})$ thresholds probably result in states with exotic properties. In this case, the situation would be similar to the $X(3872)$. If our pentaquark masses are overestimated and they are close to the masses of the $P$-wave $QQQ$ states, the configuration mixing is also possible. In this case, relatively lower $QQQ$-like states might exist. To identify the nature of such states needs information from production or decay properties. We have to wait for future experimental searches for possible interesting phenomena in triply heavy baryon systems.

\section{SUMMARY}\label{sec5}

In this paper, we have investigated the mass splittings for the pentaquark states with the configuration $QQQq\bar{q}$ in a chromomagnetic model, where $QQQ$ can be color-octet $ccc$, $bbb$, $ccb$, or $bbc$. The values of their masses cannot be determined accurately since the model does not involve dynamics. We estimate roughly the masses with two methods. Since no symmetry constraints suppress the $S$-wave rearrangement decays into $(QQq)+(Q\bar{q})$, whether the decays happen or not depends only on kinematics. Although the obtained results have uncertainties, it seems that all the studied pentaquarks are above the respective lowest $(QQq)$-$(Q\bar{q})$ type thresholds and thus do not have low masses and narrow widths. On the other hand, the proposed molecules around the thresholds of $\Xi_{cc}D$, $\Xi_{cc}D^*$, $\Xi_{cc}\bar{B}$, and $\Xi_{cc}\bar{B}^*$ are still possible \cite{Chen:2017jjn}. Based on our results, once a state around the threshold of $\Xi_{cc}D$ were observed, its molecular nature other than compact multiquark nature is favored.

To see the important mixing effects from different color-spin structures, we express the mass of an obtained pentaquark state with effective chromomagnetic interactions by varying coupling constants in the model Hamiltonian. The relevant coefficients $K_{ij}$'s are tabulated in tables \ref{varQQQqqbar} and \ref{varQQQpqqbar}. From these tables, it is easy to understand the contributions to the pentaquark mass from each pair of quark interactions. When comparing with the CMI matrices in Sec. \ref{sec3}, one finds that the mixing may change the interaction strengths significantly (even signs). Actually, the value of $K_{ij}$ can also be obtained with the eigenvectors in tables \ref{penta-massccc}-\ref{penta-massbbc} and the CMI matrices in Sec. \ref{sec3}. With the explicit expressions, one can see how the mixing effects contribute to a state. For example, the effective $cn$ interaction in the lower $J=1/2$ $cccn\bar{n}$ state ($I=1$ or 0) reads \begin{eqnarray}
K_{cn}C_{cn}=0.52^2\times0+(-0.85)^2\times(-\frac{20}{3})C_{cn}+0.52\times(-0.85)\times\frac{10}{\sqrt3}C_{cn}\times2=-9.92C_{cn},
\end{eqnarray}
which equals to $-9.99C_{cn}$ if the error bars in the eigenvector are included.

As a byproduct of the pentaquark study, we have obtained a conjecture for the mass inequalities for conventional doubly heavy baryons and conventional triply heavy baryons. Such relations may be tested in the future experimental results.

We hope the present study may stimulate further investigations about properties of conventional doubly or triply heavy baryons and multiquark states on both the theoretical side and the experimental side. Search for the exotic triply heavy baryons can be performed at LHC.

\section*{Acknowledgements}

This project is supported by National Natural Science Foundation of China (Grant Nos. 11775130, 11775132, 11635009, 11325525, 11875179) and the Natural Science Foundation of Shandong Province (Grant No. ZR2017MA002).


\end{document}